\documentclass[twocolumn]{aastex631}
\usepackage{amsmath}
\usepackage{float}
\usepackage{graphicx}
\usepackage{booktabs}

\begin{document}

\title{Discovery of 79 $\delta$ Scuti Stars in NGC 3532 Suggests a Decrease of Pulsator Occurrence with Age}

\correspondingauthor{Ian Berry}
\email{ianberry@hawaii.edu}

\author[0009-0009-3641-4137]{Ian Berry}
\affiliation{Institute for Astronomy, University of Hawai`i, 2680 Woodlawn Drive, Honolulu, HI 96822, USA}

\author[0000-0001-8832-4488]{Daniel Huber}
\affiliation{Institute for Astronomy, University of Hawai`i, 2680 Woodlawn Drive, Honolulu, HI 96822, USA}

\author{Yaguang Li}
\affiliation{Institute for Astronomy, University of Hawai`i, 2680 Woodlawn Drive, Honolulu, HI 96822, USA}

\author{Daniel Hey}
\affiliation{Institute for Astronomy, University of Hawai`i, 2680 Woodlawn Drive, Honolulu, HI 96822, USA}

\author[0000-0001-5222-4661]{Timothy R. Bedding}
\affiliation{Sydney Institute for Astronomy, School of Physics, University of Sydney, NSW 2006, Australia}

\author[0000-0002-5648-3107]{Simon J. Murphy}
\affiliation{Centre for Astrophysics, University of Southern Queensland, Toowoomba, QLD 4350, Australia
}

\begin{abstract}
    Many A–F type stars do not display $\delta$ Scuti pulsations, despite being located within the instability strip. Open clusters provide a unique opportunity to study $\delta$ Scuti pulsations among coeval populations with uniform chemical composition. Here we use data from the TESS Mission to discover 79 $\delta$ Scuti pulsators in the 300 Myr old open cluster NGC 3532, the largest number found within a single open cluster to-date. We report a 50$\pm$5\% pulsator fraction in NGC 3532, considerably lower than in younger stellar populations, such the Pleiades (110 Myr), NGC 2516 (100 Myr), and the Cep-Her Complex ($\leq\,$80 Myr), and similar to the pulsator fraction found among field star samples. We introduce the concept of pulsator occurrence, which corrects for incompleteness, and find it to be 63$\pm6\%$. For the stars that do pulsate, we find that the hotter stars occupy a distinct branch in the color-magnitude diagram (CMD) due to faster rotation ($>\,$150 km/s) than their non-pulsating counterparts. These results suggest that pulsator occurrence decreases with age and that rapid rotation is important in maintaining $\delta$ Scuti pulsations over time. We also investigate the Period-Luminosity (P-L) relation and the $\nu_{\rm max}$--$T_{\rm eff}$ relation of $\delta$ Scuti stars in NGC 3532. We find much scatter in the P-L relation of the dominant mode and two distinct branches in the $\nu_{\rm max}$--$T_{\rm eff}$ relation, similar to the Cep-Her Complex.

\end{abstract}

\section{Introduction}

The $\delta$ Scuti pulsators are spectral type A0-F5 stars on or near the main sequence within the classical Cepheid instability strip, spanning effective temperatures ($T_{\rm eff}$) between $\sim7500$-$9500$ K. These stars are known to display high-frequency pressure (p) modes due to internal helium ionization zones, which drive pulsations through cyclical changes in opacity known as the $\kappa$ mechanism \citep{2004A&A...414L..17D}. Some instability strip stars are hybrid pulsators, displaying both $\delta$ Scuti pulsations and low-frequency gravity (g) modes associated with the $\gamma$ Doradus pulsators \citep{2010AN....331..989G, 2010arXiv1007.3176H,2014MNRAS.444..102K,2018FrASS...5...43B}.

The physics of the driving mechanism of pulsations in $\delta$ Scuti stars remains incompletely understood despite the decades of observations and modeling since the discovery of the prototype \citep{1900ApJ....12..254C}. One lingering issue is that some stars that lie within the instability strip seem not to pulsate at all, contrary to theoretical expectations \citep{2004A&A...414L..17D,2005A&A...435..927D}. The opposite scenario also seems to be true in certain cases, i.e. stars that appear to fall outside the bounds of the instability strip can show $\delta$ Scuti pulsations \citep[e.g.][]{2011A&A...534A.125U,2018FrASS...5...43B,2019MNRAS.485.2380M,2023A&A...674A..36G}.

\begin{figure*}
    \centering
    \includegraphics[width=\linewidth]{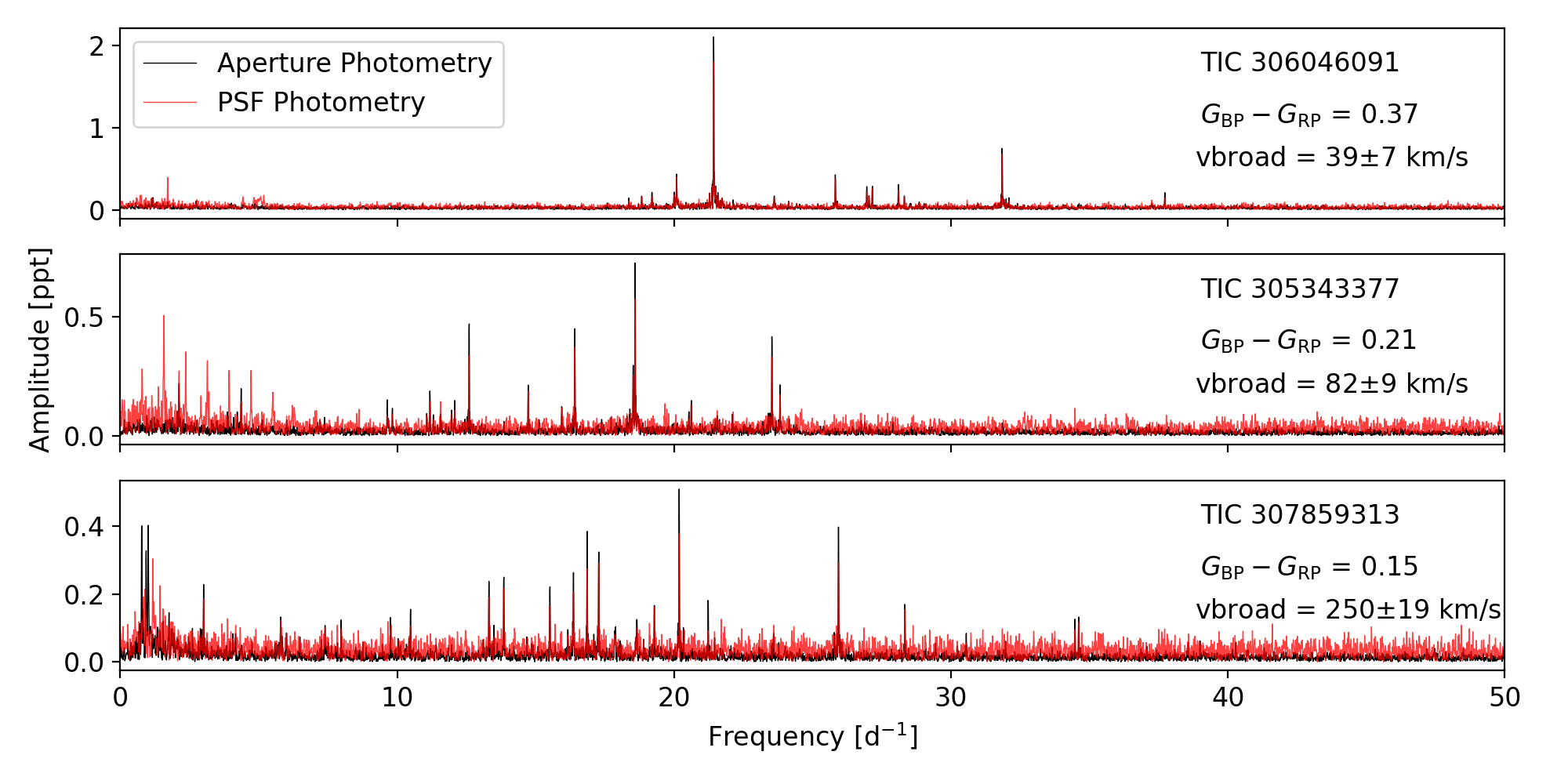}
    \caption{Example amplitude spectra of three $\delta$ Scuti stars in NGC 3532, from aperture photometry (black) and PSF photometry (red). The stars are ordered by Gaia \texttt{vbroad}, and the TIC IDs and Gaia $G_{\rm BP}-G_{\rm RP}$ are provided.}
    \label{fig:sample_pgs}
\end{figure*}

This has led to much work in the determination of a $\delta$ Scuti ``pulsator fraction'', the ratio of pulsators to the total number of stars in a sample, which appears to vary between different stellar populations/samples. Among field stars, \citet{2019MNRAS.485.2380M} used \textit{Kepler} to measure a maximum 70\% pulsator fraction.  \cite{2024MNRAS.528.2464R} used the Transiting Exoplanet Survey Satellite \citep[TESS;][]{2015JATIS...1a4003R} to observe a large sample of field $\delta$ Scuti stars within a narrow color range, and found a higher pulsator fraction among brighter stars compared to fainter stars. This drop in pulsator fraction suggested that detection capabilites/biases may play some part in whether an instability strip star shows $\delta$ Scuti pulsations. Similar results were found by \cite{2025arXiv250818589M} using a sample of bright TESS stars spanning the width of the instability strip.

Physical factors must also govern whether a star within the instability strip pulsates, and these may vary with age. $\delta$ Scuti pulsations require the presence of a helium ionization zone at a specific depth and temperature ($\sim 10,000\,{\rm K}$) in the star in order to drive pulsations through the $\kappa$ mechanism. In slowly rotating stars, helium can sink out of the ionization zone over time \citep{1973A&A....23..221B}, stopping pulsations from the $\kappa$ mechanism. For a 1.6 ${\rm M_\odot}$ star with a stellar wind, 50\% of helium will be depleted from the ionization zone by 100 Myr, and up to 80\% by 500 Myr \citep{2005A&A...443..627T,2016A&A...589A.140D}. Therefore, the youngest instability strip stars should still have the necessary helium in the ionization zone to drive pulsations, regardless of other stellar parameters.

\begin{figure*}
    \centering
    \includegraphics[width=.95\linewidth]{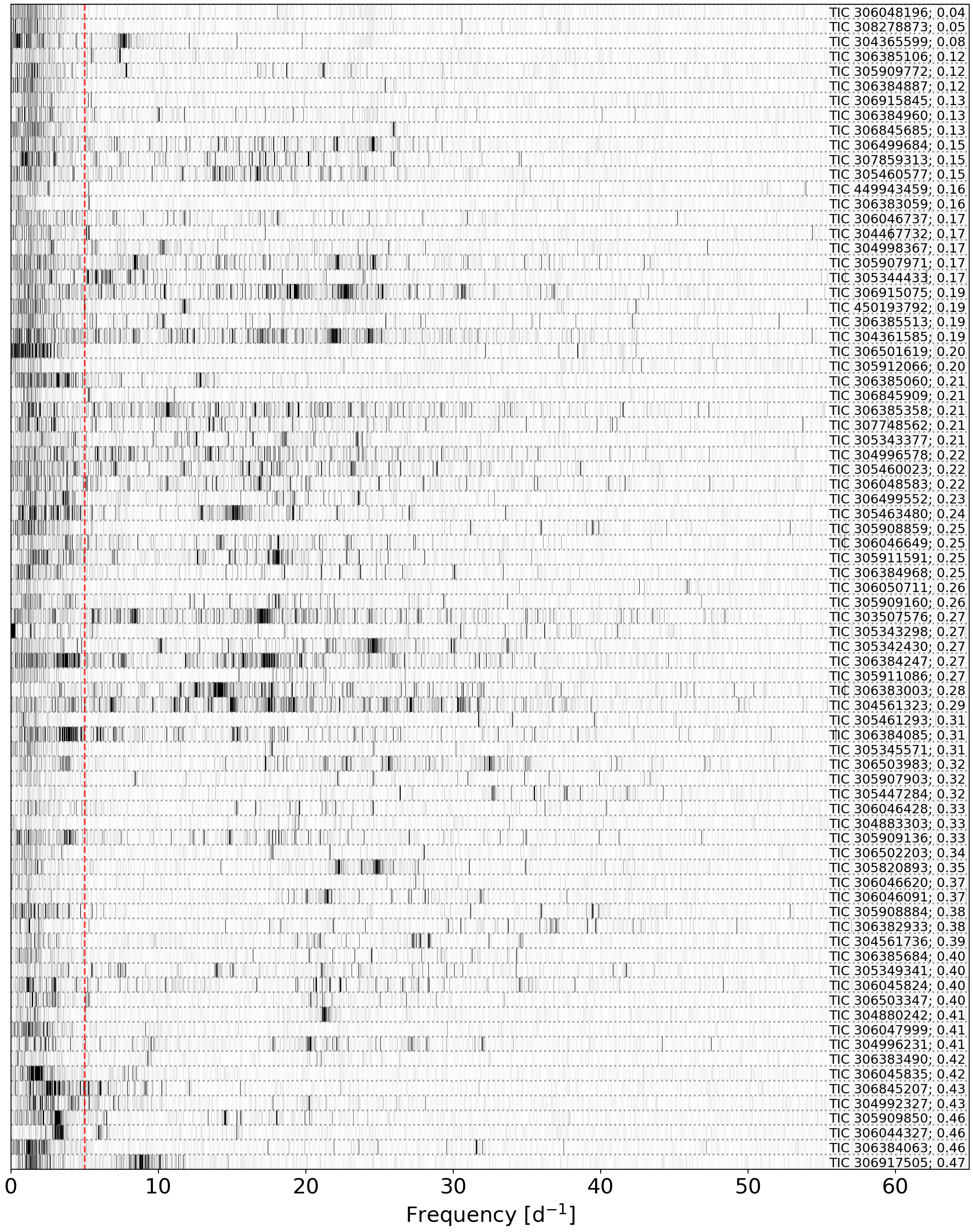}
    \caption{Stacked amplitude spectra of all 79 $\delta$ Scuti stars in NGC 3532. Darker colors denote larger amplitudes. Stars are ordered by Gaia $G_{\rm BP}-G_{\rm RP}$, with bluest stars on top. The dashed red line marks 5 ${\rm d^{-1}}$, which is commonly used as the boundary between $\delta$ Scuti pulsations and other sources of low frequency variability \citep[e.g][]{2002MNRAS.333..251H, 2010arXiv1007.3176H,2019MNRAS.485.2380M}. Each amplitude spectrum is labeled with its corresponding star and color (without dereddening).}
    \label{fig:stacked_pg}
\end{figure*}

Another astrophysical process affecting the driving of pulsations is rotation. Gravitational diffusion of helium can be easily disrupted by rotational mixing through rapid rotation \citep{2004A&A...425..591H}. Recently, \citet{2024ApJ...972..137G} found a clear increase in pulsator fraction with increasing rotation rate using a large sample of $\delta$ Scuti pulsators. Similar results were found by \cite{2024MNRAS.534.3022M} with $\delta$ Scuti stars in the Cep-Her Complex. Therefore, we may expect to see a relatively high pulsator fraction even in older stellar populations, but we would also expect to see such pulsators as rapid rotators. Such a scenario could be possible as stars above the Kraft Break ($T_{\rm eff}\gtrsim6200$ K) typically do not spin-down significantly during the main sequence \citep{1967ApJ...150..551K}.

Finally, chemical composition can affect pulsator occurrence. Chemically peculiar Am stars, for example, display strong metal absorption lines. These stars are known to generally be slow rotators, which is thought to allow helium to gravitationally diffuse out of the ionization zone, while allowing for metals to rise to the surface through radiation pressure \citep{1974NInfo..32..104P, 1980LNP...125...22D, 2015A&A...579A.116O}. At the same time, these stars appear to have lower pulsation fractions than their non-metallic-lined counterparts \citep{1970ApJ...162..597B, 2021arXiv210709479G}. Am stars with $\delta$ Scuti pulsations are known to exist, and the pulsator fraction is indeed low \citep[$\sim13.5\%$;][]{2024A&A...690A.104D}.

Open clusters offer an ideal environment for measuring pulsator fractions at a fixed age and chemical composition. Pulsator fractions have been measured with TESS in three young stellar populations: the Pleiades \citep[100 Myr;][]{2023ApJ...946L..10B}, NGC 2516 \citep[100 Myr;][]{2024A&A...686A.142L} and the Cep-Her complex \citep[$\leq$ 80 Myr;][]{2024MNRAS.534.3022M}, where maximum pulsator fractions were measured to be $\sim$80\%, $\sim$80\% and 100\%, respectively. These higher pulsation fractions suggest that the pulsation fraction depends on age. However, pulsation fractions in older, intermediate-age populations have yet to be measured.

The NASA TESS mission \citep{2015JATIS...1a4003R,2024arXiv241012905W} allows for systematic study of $\delta$ Scuti stars in open clusters. During the Prime mission of TESS, full-frame images (FFIs) were downloaded every 30 minutes, limiting the study of $\delta$ Scuti stars because many pulsate near or above the 24 cycle ${\rm d}^{-1}$ Nyquist frequency. In the TESS extended missions, the FFI cadence was reduced to 10 minutes and then to 200 seconds, which is fast enough to identify almost all $\delta$ Scuti pulsators, regardless of their pulsation frequencies. These FFIs now cover more than 90\% of the sky, meaning almost all nearby ($\lesssim$ 500 pc) $\delta$ Scuti stars can be studied, both in the field and within open clusters.

\begin{figure}
    \centering
    \includegraphics[width=\columnwidth]{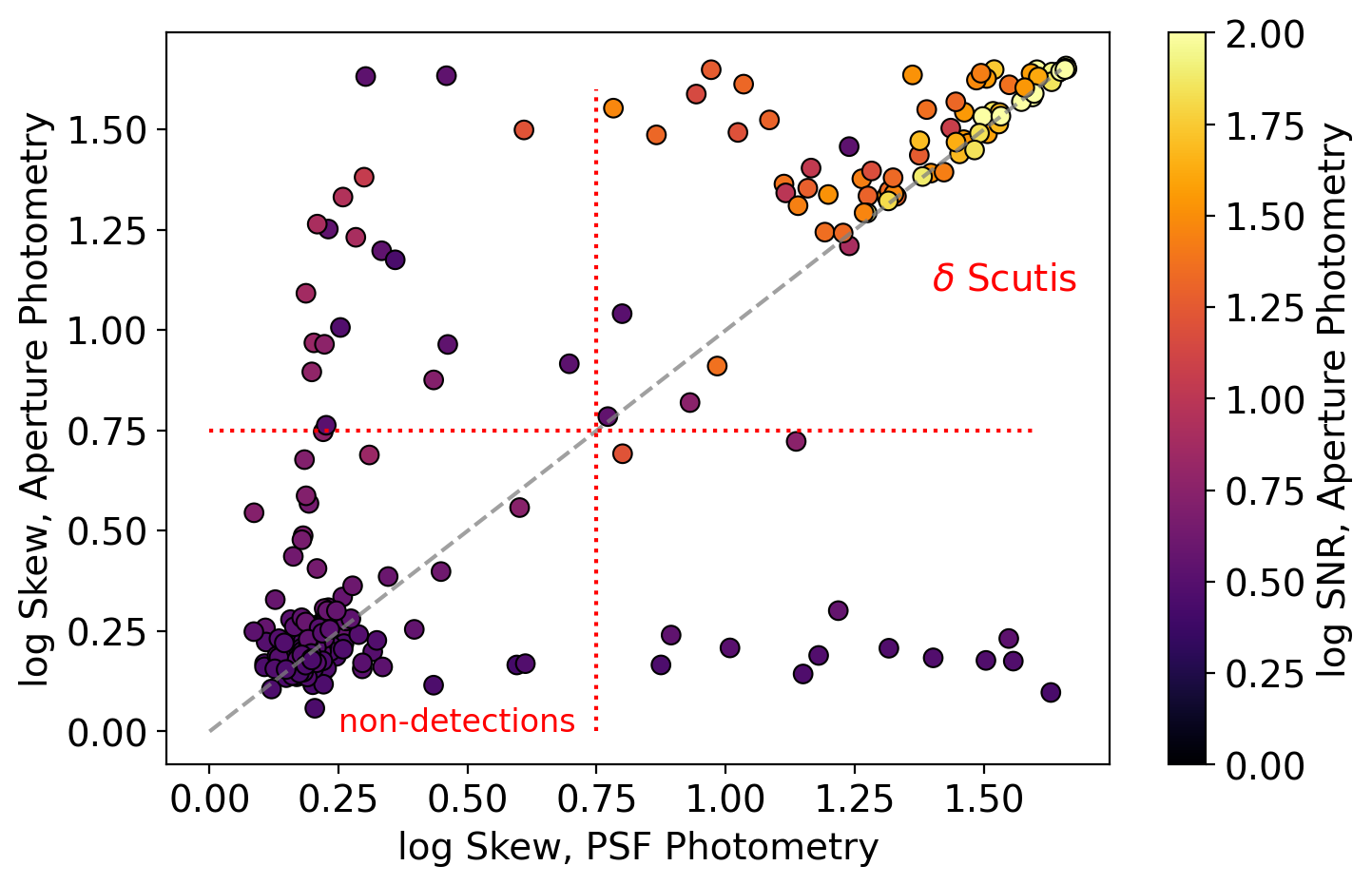}
    \caption{Scatter plot showing the log skewness of peak amplitudes in aperture and PSF photometry from TGLC. We apply two lower boundaries in log skewness, 0.75 for both aperture and PSF photometry. Points are color-coded by the signal-to-noise ratio (SNR) of the highest peak in the periodogram. The gray dashed line is the 1:1 line. }
    \label{fig:sct_identification}
\end{figure}

In this paper, we use TESS photometry to identify and characterize 79 $\delta$ Scuti stars in the $\sim$300 Myr old cluster NGC 3532 \citep{2011AJ....141..115C,2012A&A...541A..41M,2022MNRAS.509.1912O}, the most found within an individual open cluster to-date. We use this sample to investigate the relationship between rotation and $\delta$ Scuti pulsations (\S\ref{sec:rot}), measure the pulsator fraction and introduce pulsator occurrence (\S\ref{sec:frac}), and analyze the Period-Luminosity (\S\ref{sec:P_L}) and $\nu_{\rm max}$--$T_{\rm eff}$ (\S\ref{sec:vmax_Teff}) relations. In addition, we measure the large frequency separation $\Delta\nu$ for one $\delta$ Scuti star (\S\ref{sec:Dnu}).

\section{Data and Methods} \label{sec:methods}

\subsection{NGC 3532 Membership}

To find NGC 3532 cluster members we used a catalog provided by \citet{2023A&A...673A.114H,2024A&A...686A..42H}, who conducted a blind census of Galactic open clusters using Gaia DR3 \citep{2023Gaia} astrometry, and the Hierarchical Density-Based Spatial Clustering of Applications with Noise \citep[HDBSCAN;][]{mcinnes2017hdbscan} clustering algorithm. \citet{2023A&A...673A.114H} provided membership probabilities and a flag indicating whether a member candidate falls within an estimated tidal radius of the cluster, calculated using a definition from \cite{1962AJ.....67..471K}. We only searched for $\delta$ Scuti pulsators among candidate members where this flag is true. By design, all stars that fall within the tidal radius have membership probability $>50\%$. \citet{2023A&A...673A.114H} noted that stars with membership probability $<50\%$ always correspond to low quality candidate members. 1842 candidate members fall within the estimated tidal radius of NGC 3532 and are considered good members.

\subsection{Interstellar Reddening}

We corrected for interstellar reddening in the Gaia photometry, first adopting $E(B-V)=0.034$ from \cite{2019A&A...622A.110F}. We then estimated $E(G_{\rm BP}-G_{\rm RP})=0.044$ using the relation $E(B-V)=0.76\,E(G_{\rm BP}-G_{\rm RP})$, which is calculated from Table~2 in \citet{2019ApJ...877..116W}.

\subsection{TESS Photometry}

We used TESS FFI photometry to search for $\delta$ Scuti pulsators in NGC 3532, which was observed in Sector 37 (mission year 3, 2021 14 January to 8 February) at 10-minute cadence, and in Sectors 63 and 64 at 200-second cadence (mission year 5, 2023 10 March to 6 April, and 2023 6 April to 4 May, respectively).

NGC 3532 (C1104$-$584 in the IAU nomenclature) is a populous open cluster and its Galactic coordinates ($l=289.5^\circ$, $b=1.4^\circ$) place it well within the plane of the Milky Way. Therefore, we expect that blending and contamination can affect the light curves given the large 21$\arcsec\,$ pixels of TESS. To mitigate this, we used TESS-Gaia Light Curves \citep[TGLCs;][]{2023AJ....165...71H}, which models the TESS FFIs using the effective point-spread function and uses Gaia DR3 astrometry and photometry as fixed priors to reduce contamination from nearby sources. 

We produced the TGLCs for each star and each sector using the \texttt{quick\textunderscore lc.tglc\textunderscore lc} method from the \textsc{tglc} Python package. This method produces both calibrated aperture and PSF light curves. We left most settings to the default, including the $3\times3$ pixel aperture size. However, we set the FFI cutout size to $55\times55$ pixels to improve computation time (default is 90$\times$90).

\begin{figure*}
    \centering
    \includegraphics[width=\linewidth]{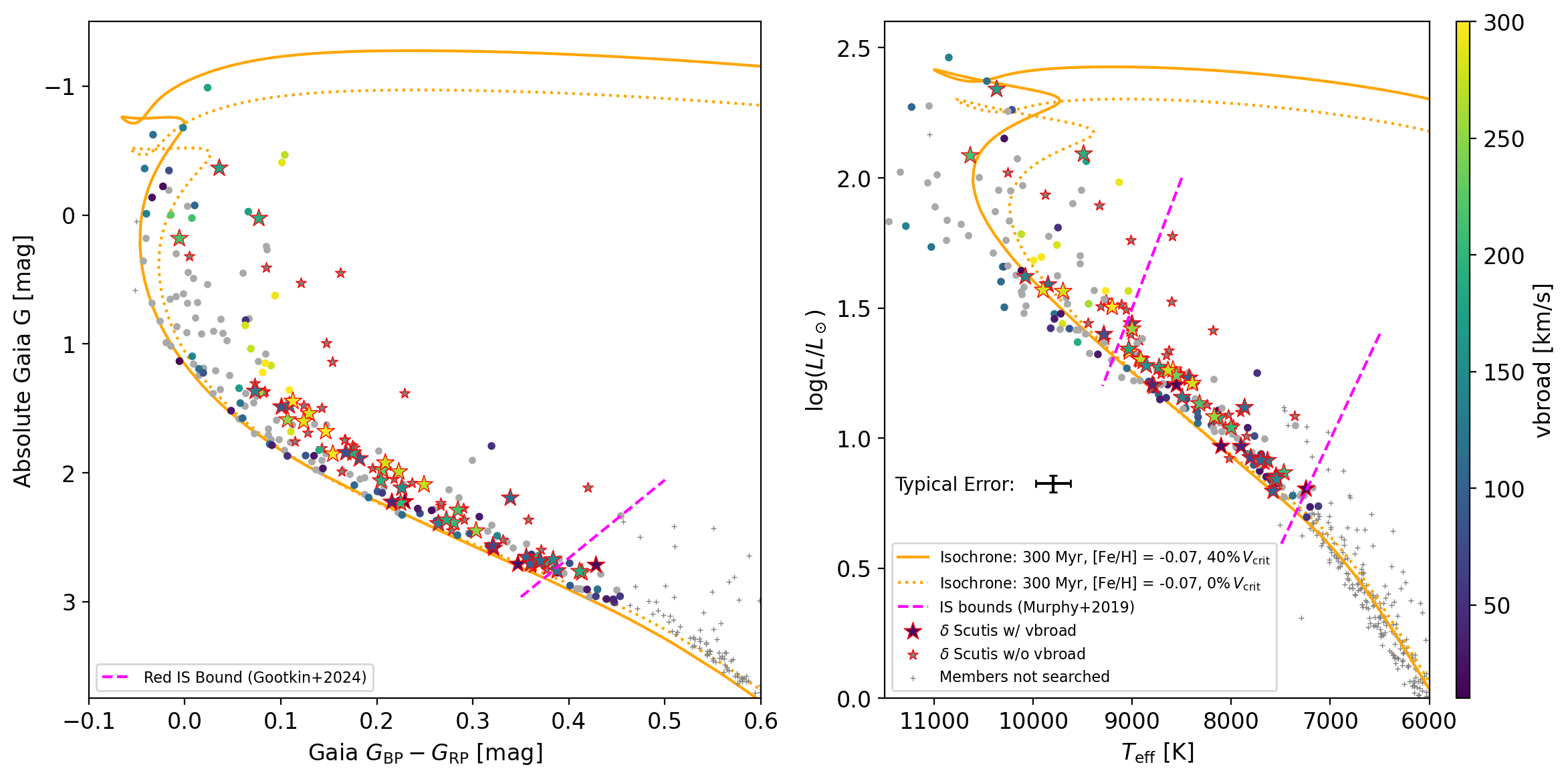}
    \caption{\textit{Left}: Dereddened Gaia Color-Magnitude Diagram (CMD) of NGC 3532. Color-mapped points are stars with a Gaia \texttt{vbroad} measurement, and gray points have no \texttt{vbroad} measurement. Star markers with red outlines are $\delta$ Scuti pulsators, while circular points are non-pulsators. Gray pluses are stars we did not search. Furthermore we plot MIST isochrones with age 300 Myr and [Fe/H] = -0.07 dex. The dotted isochrone is a non-rotating model, and the solid isochrone is a rotating model at 40\% critical rotation speed. An empirically derived red instability strip (IS) edge from \cite{2024ApJ...972..137G} is shown as the dashed magenta line. \textit{Right}: As Left except in $\log L$ vs. $T_{\rm eff}$ space. Luminosities and $T_{\rm eff}$ values come from the TIC catalog \citep{2019AJ....158..138S}. The magenta lines now show the empirically derived IS bounds from \cite{2019MNRAS.485.2380M}.}
    \label{fig:cmd}
\end{figure*}

We first produced light curves in each TESS Sector for cluster members with $G_{\rm BP}-G_{\rm RP} = [0, 0.5]$, which corresponds to 247 stars total. For each star, the TGLCs from each sector were stitched together, and an amplitude spectrum was calculated using the \textsc{AstroPy} Python package \citep{astropy:2013,astropy:2018,astropy:2022}. This was done on both the aperture and PSF light curves. Amplitude spectra for three $\delta$ Scuti stars in this cluster are shown in Figure \ref{fig:sample_pgs}, and a stacked amplitude spectrum for all stars in the cluster is shown in Figure \ref{fig:stacked_pg}. In general, the aperture photometry provides a higher signal-to-noise ratio (SNR) than PSF photometry. Also, the PSF amplitudes tend to be smaller than those in the aperture photometry. This is a known characteristic of the TGLCs, and arises from imperfect decontamination \citep{2023AJ....165...71H}. Here, the PSF photometry is used only if a nearby A or F-star cluster member exists within 4 TESS pixels from the source, in order to mitigate the effects of blending (discussed in \S\ref{subsec:blend}).

\subsection{$\delta$ Scuti Identification}

\subsubsection{Methodology}

To identify $\delta$ Scuti pulsators, we followed \citet{2019MNRAS.485.2380M} by using the skewness of peak amplitudes, which will be systematically larger for $\delta$ Scuti stars than for stars with no high-frequency variability. To avoid false positives from sources of low-frequency variability, e.g. rotational modulation or $\gamma$ Doradus (Dor) pulsations \citep{Kaye_1999}, we only considered peaks with frequencies $>$7 ${\rm d^{-1}}$ when calculating skewness.

Figure \ref{fig:sct_identification} shows the skewness of peak heights using the amplitude spectra from both the aperture and PSF light curves. We visually define a lower bound on log skewness of 0.75 for both aperture and PSF photometry to automatically identify the $\delta$ Scuti stars. Initially, 92 stars are identified as $\delta$ Scuti pulsators with this method.

For some stars the skewness is high in aperture photometry, but low in PSF photometry. An explanation for this is blending/contamination in the $3\times3$ pixel aperture from nearby $\delta$ Scuti sources that is removed or reduced in the PSF photometry. Another reason could be that some stars show lower amplitude pulsations that appear in the aperture photometry, but are lost in the noisier PSF photometry. Twelve stars show high skewness from the PSF photometry, and low skewness from the aperture photometry. The amplitude spectra of these stars show one or more low frequency peaks with harmonics that exist above 7 ${\rm d^{-1}}$. This could possibly be the result of a poor fit when modeling the PSF, and these stars are not classified as $\delta$ Scuti pulsators.

\begin{figure*}
    \centering
    \includegraphics[width=\linewidth]{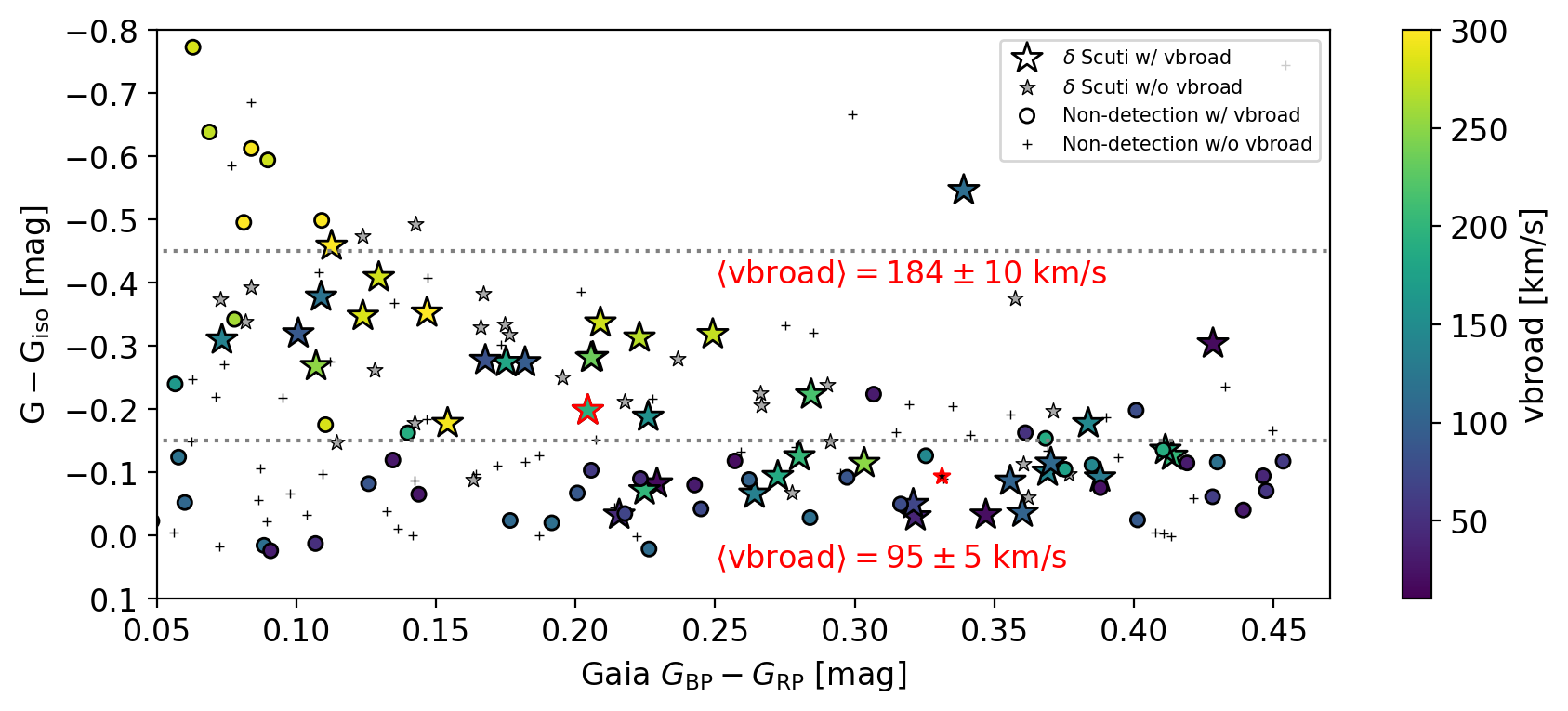}
    \caption{Vertical distance above the 40\% critical isochrone (solid orange curve in the left panel of Figure \ref{fig:cmd}) vs. $G_{\rm BP}-G_{\rm RP}$. Color-mapped points are stars with a \texttt{vbroad} measurement. Star points are $\delta$ Scuti stars, and circular points and gray pluses are non-pulsators. Stars with red edges are $\delta$ Scuti stars that suffer from blending (described in \S \ref{subsec:blend}). This scatter plot shows that the $\delta$ Scuti stars fall farther from the isochrone and rotate more rapidly than the non-pulsators on the lower branch. This is most apparent at $G_{\rm BP}-G_{\rm RP}\lesssim 0.25$. To show this more explicitly, we found the mean \texttt{vbroad} of stars on the upper branch, designated as the region between the two gray dotted lines, where $\langle\texttt{vbroad}\rangle\approx184$ km/s. This is about twice the rotation seen on the lower branch, where $\langle\texttt{vbroad}\rangle\approx95$ km/s. Roughly two-thirds of the $\delta$ Scuti stars in NGC 3532 fall on the upper branch.  }
    \label{fig:dGiso}
\end{figure*}

\subsubsection{False Positives}

The largest source of false positives in our $\delta$ Scuti sample is eclipsing binaries (EBs), which we identified following \cite{2024ApJ...972..137G} by using the number of points above the mean flux in the light curve. Because the variability of $\delta$ Scuti stars is approximately sinusoidal, the number of points above the mean flux level should be $\approx 50\%$. Meanwhile, the light curves of EBs are dominated by deep eclipses, so the number of points above the mean flux level will be $<50\%$. We adopt a threshold of 45\% to identify EBs. Three stars initially classified as $\delta$ Scuti stars are found to be EBs and were removed from the sample. Two of these EBs are GV Car and HD 303734 (TIC 306047507 and 305343777, respectively) and were discovered in previous literature \citep{2006Ap&SS.304..199S,2022MNRAS.509.1912O}. The third EB is TIC 306390348.

\subsubsection{Contamination/Blending} \label{subsec:blend}

Given the crowded field around NGC 3532, and the large 21$\arcsec\,$ pixels of TESS, blending of sources is to be expected. High-amplitude pulsations in $\delta$ Scuti stars, combined with the fact that the brightest stars in the field are cluster members, make contamination from fainter stars negligible. Therefore, the most relevant contamination scenario involves two blended A- or F-type cluster members where one or both stars are pulsating, making it difficult to disentangle which pulsation modes belong to which star. To identify blended pairs of A/F star members, we calculated the nearest neighbor distance for all 247 member A/F stars, and found the Pearson Correlation Coefficient on the PSF amplitude spectra for all pairs of stars. We found one case of a blended pair which has Pearson r $>0.5$ and a nearest neighbor distance about 12$\arcsec$ apart. These two stars are TIC 305908895 and 305908884 and are both included in this analysis as $\delta$ Scuti pulsators. 

\subsubsection{Photometric Quality Cuts}

Some stars have one or more TESS sectors with poor photometry. To identify these sectors, we compared the root-mean-square (RMS) scatter of each observed light curve to that of a purely noise-driven model. The model RMS was computed using the expected TESS noise levels derived using the \textsc{ticgen} Python package \citep{jaffe_barclay_2017,2018AJ....156..102S}. For all sectors and stars, we calculated the ratio of the observed RMS to the model RMS and defined the upper limit as the $1\sigma$ deviation above the mean of the resulting distribution. We find RMS upper limits for the 10-minute and 200-second cadence sectors to be factors of 10 and 5.5 above the model RMS, respectively. To avoid confusing genuine variability with poor photometry, we also compared the noise levels in the amplitude spectra with those from the model light curves, and set an upper limit of five times the model noise level.  This method does not hinder our ability to detect $\delta$ Scuti pulsators, and a vast majority of the $\delta$ Scuti stars have no bad sectors of photometry; only four have a single bad sector and one has two bad sectors. Non-detections with bad photometry in two or three TESS sectors were not considered when computing pulsator fraction and occurrence in \S\ref{sec:frac}, as these stars cannot be reliably classified.

\subsection{Final Catalog}
In total, we identify 79 $\delta$ Scuti pulsators in NGC 3532. The properties of all 247 stars we searched are provided in Table \ref{tab:delta_scuti}.

\begin{figure}
    \centering
    \includegraphics[width=\columnwidth]{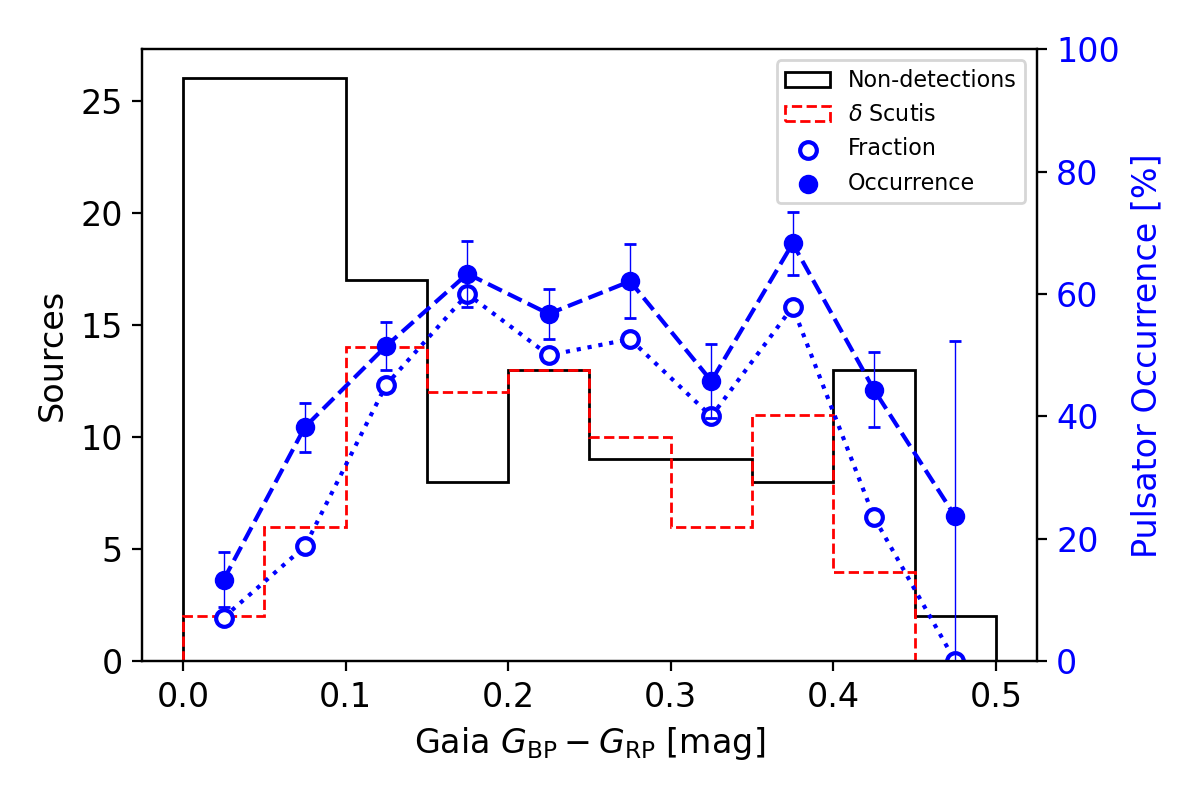}
    \caption{Histograms, with numbers on the left axis, showing the population distribution of $\delta$ Scuti stars (dashed red) and non-$\delta$ Scuti stars (black) as a function of $G_{\rm BP}-G_{\rm RP}$. The pulsator fraction in each bin is overplotted as open circles connected by the dotted blue line, with values on the right axis. The pulsator occurrence is shown as the blue dashed line and points points. The pulsator fraction is heavily dependent on color, and peaks at 60$\pm$11\%. The pulsator occurrence peaks at 67$\pm5\%$. Across the center of the instability strip, ($G_{\rm BP}-G_{\rm RP}=[0.15,0.4]$) the pulsator fraction and occurrence is 50$\pm5\%$ and 59$\pm5\%$, respectively.}
    \label{fig:pulse_frac}
\end{figure}

\section{$\delta$ Scuti Pulsations and Rotation} \label{sec:rot}

Figure \ref{fig:cmd} shows the location of pulsators and non-detections in the Gaia CMD and the HR diagram. 
We observe two clear branches in the CMD, with the upper branch populated by $\delta$ Scuti pulsators, and the lower branch populated by non-pulsators. This feature becomes more prominent among the hotter stars.

This feature appears to be due to rapid rotation, which is known to broaden the main sequence \citep{1999AcA....49..119P,2011A&A...533A..43E}, due to gravity darkening. Our results are also consistent with the idea that the pulsator fraction among main sequence stars in the instability strip generally increases with more rapid rotation \citep{2024ApJ...972..137G}. To test this further, we obtained Gaia \texttt{vbroad} measurements, which measures spectral line broadening using the Gaia Radial Velocity Spectrometer \citep{2018A&A...616A...5C,2022gdr3.reptE...6S, 2023A&A...674A...8F}. This measurement includes \textit{all} sources of line broadening in the calcium triplet at 846--870 nm, including rotation. Comparisons with independently measured $v\sin i$ show that \texttt{vbroad} agrees well for rotation velocities $75\leq v\sin i\leq 200$ km/s, but tends to overestimate very rapid rotators ($v \sin i > 200$ km/s), and tends to underestimate slow rotators ($v \sin i < 50$ km/s) by 10--15\% \citep{2023A&A...674A...8F,2024MNRAS.534.3022M}.

Figure \ref{fig:cmd} shows that the $\delta$ Scuti pulsators on the upper branch of the Gaia CMD indeed have higher \texttt{vbroad} values than the non-detections on the lower branch. To display this more explicitly, Figure \ref{fig:dGiso} shows the vertical height above the 40\% critical rotation isochrone, which we label as $G-G_{\rm iso}$, as a function of color. The two branches, ending at $G_{\rm BP}-G_{\rm RP} \approx 0.3$, are clearer and confirm that $\delta$ Scuti pulsators inhabit the upper branch and rotate more rapidly than the non-detections on the lower branch. If we define the upper branch as the region with  $-0.45 \leq G-G_{\rm iso} \leq -0.15$, and the lower branch with $G-G_{\rm iso} > -0.15$, then we find that the mean \texttt{vbroad} of stars within the upper branch is $184\pm10$ km/s, and is $95\pm5$ km/s within the lower branch. Furthermore, the upper branch contains 54 $\delta$ Scuti stars, just over two-thirds of the total $\delta$ Scuti population in this cluster. These results provide further evidence that rotation is a highly important factor to consider when studying $\delta$ Scuti stars.

Seven of the $\delta$ Scuti stars in this cluster have been found to show ${\rm H}\alpha$ emission \citep[TIC 306044327, 306385684, 306045824, 306046091, 306047999, 305909850, 304996231;][]{2025ApJ...979..246H}. All but one of these stars have $\texttt{vbroad}>100$ km/s, consistent with the independently measured $v\sin i$ from \cite{2025ApJ...979..246H}. One star has a slower $v\sin i\approx 40$ km/s, and might be viewed at a lower inclination. The rapid rotation suggests that the ${\rm H}\alpha$ emission from these $\delta$ Scuti stars arises from a decretion disk caused by rapid rotation, similar to that of the Be stars \citep{2024arXiv241106882R}. 

\begin{figure}
    \centering
    \includegraphics[width=\columnwidth]{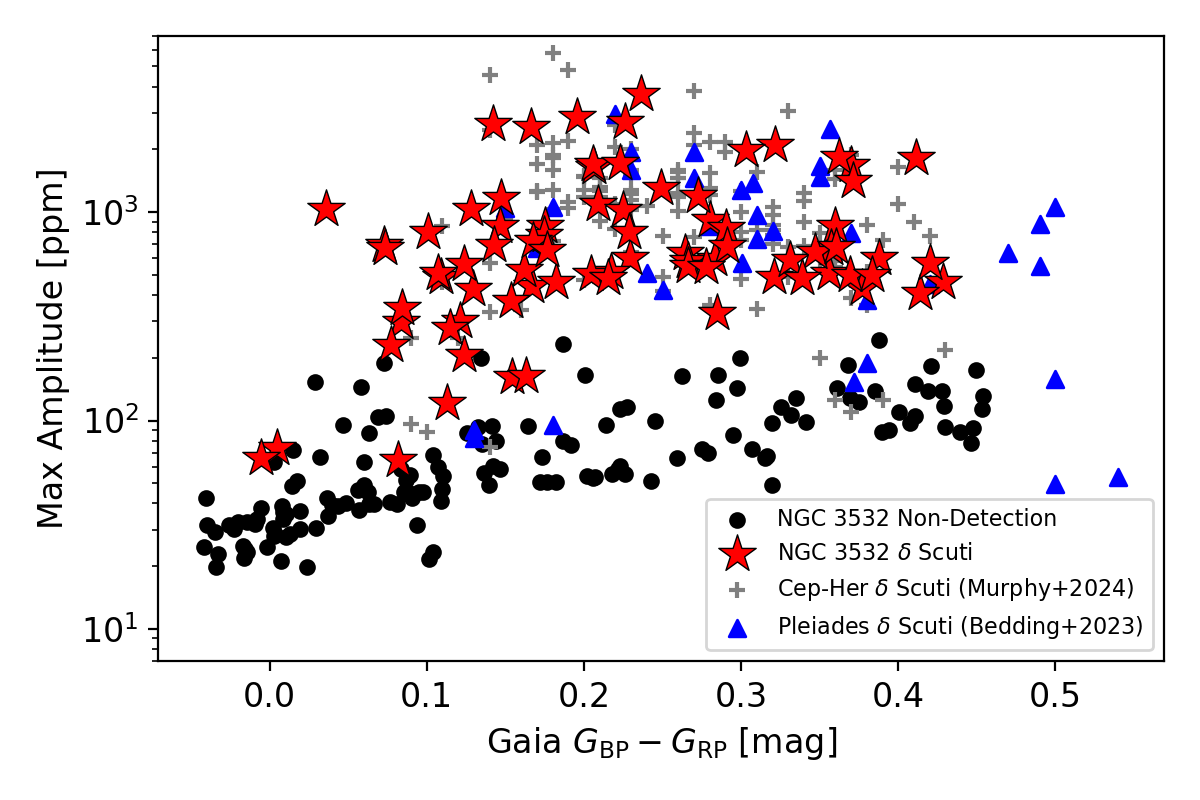}
    \caption{Maximum amplitude vs. Gaia $G_{\rm BP}-G_{\rm RP}$ for non-detections and $\delta$ Scuti stars in NGC 3532 as the black points and red stars, respectively. For comparison, $\delta$ Scuti stars in the Pleiades and the Cep-Her Complex are overplotted as the blue triangles and grey pluses, respectively.}
    \label{fig:complete}
\end{figure}

Some $\delta$ Scuti stars on the lower branch in the CMD exhibit slow rotation ($\lesssim 50$ km/s). These stars are most likely intrinsically slow rotators. If these stars were rapid rotators seen at low inclination, they would be over-luminous due to gravity brightening, and therefore would not appear so close to the isochrone. Confirming the inclination requires a rotation period, which has been measured for some A stars from rotational modulations \citep{2015A&A...577A..64B,2017MNRAS.467.1830B,2024AJ....168...13S}. However, searching for rotational modulation from $\delta$ Scuti NGC 3532 members would require a longer time series than is available from TESS.

Not all rapidly rotating $\delta$ Scuti stars are on the upper branch. One example is TIC 305343298, which is a $\delta$ Scuti pulsator with a \texttt{vbroad}$\approx 200$ km/s, yet lies within the lower branch at $G_{\rm BP}-G_{\rm RP}=0.22$. Inspection of this star's amplitude spectrum shows a high-amplitude, low-frequency peak with harmonics, consistent with an EB. This star's light curve also shows apparent shallow eclipses. Given the apparent binarity of this $\delta$ Scuti star, its position on the CMD may be inaccurate.

\section{Pulsator Fraction and Occurrence} \label{sec:frac}

Figure \ref{fig:pulse_frac} shows the pulsator fraction in NGC 3532 as a function of color (open blue cicles). This is the fraction of observed stars in which pulsations were detected, and reaches a maximum of 60$\pm$11\% at $G_{\rm BP}-G_{\rm RP}\simeq 0.2$. The pulsator fraction across the center of the instability strip ($G_{\rm BP}-G_{\rm RP}=[0.15,0.4]$) is 50$\pm5\%$.

\begin{figure}
    \centering
    \includegraphics[width=\columnwidth]{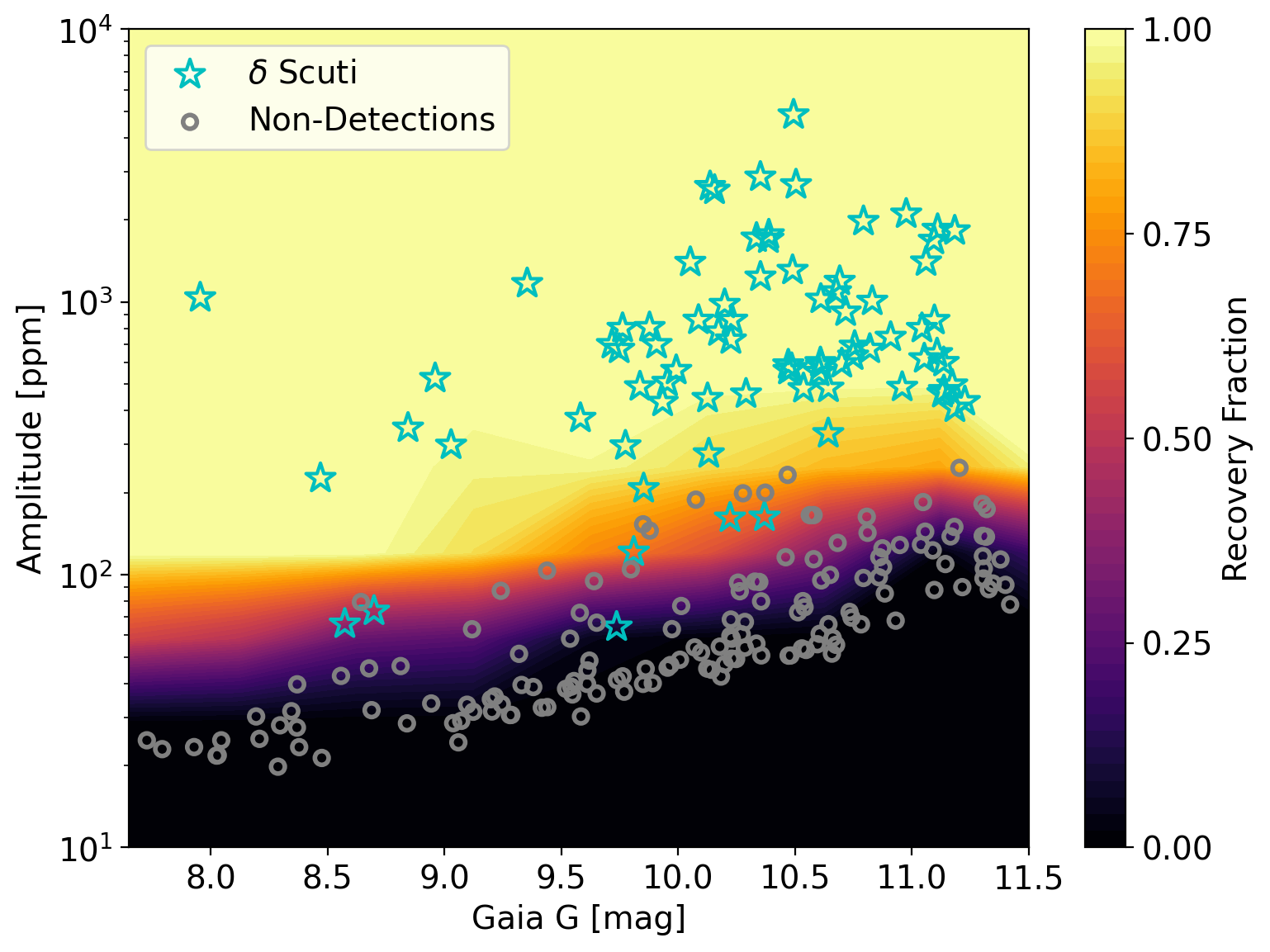}
    \caption{Recovery rates, shown as the contour plot, as a function of maximum amplitude and apparent Gaia $G$ magnitude. Non-detections are plotted as the grey circles, and $\delta$ Scuti stars are plotted as the cyan stars.}
    \label{fig:recovery}
\end{figure}

\begin{figure}
    \centering
    \includegraphics[width=\columnwidth]{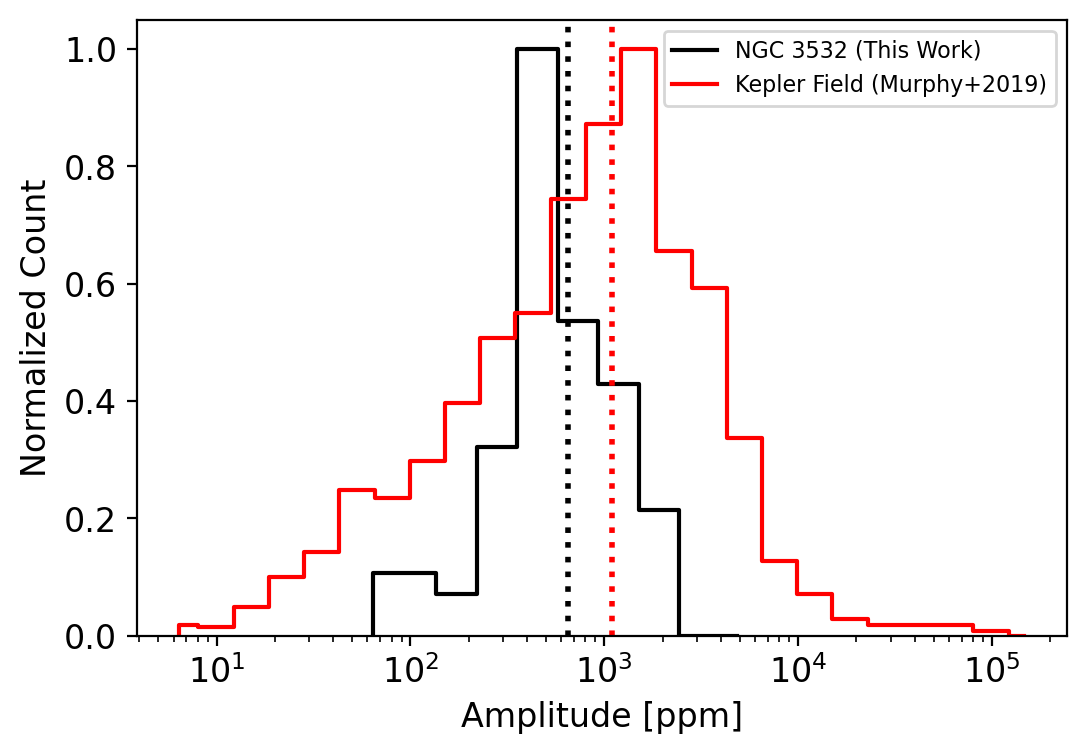}
    \caption{Distributions of maximum amplitudes for NGC 3532 (black) and the \textit{Kepler} Field \citep[red;][]{2019MNRAS.485.2380M}. The dotted lines show the median amplitude for each distribution, approximately 650 and 1100 ppm, respectively. The \textit{Kepler} amplitudes have been passband-corrected, using an amplitude ratio of 0.8 \citep{2019MNRAS.489.1072L}.}
    \label{fig:amp_dists}
\end{figure}

The mean Gaia magnitude for our sample is $G \approx 10$. The asteroseismic detection capabilities of TESS decrease significantly above this magnitude \citep{2024MNRAS.528.2464R}. Therefore, rather than considering pulsator fraction, it may be more appropriate to infer the pulsator \textit{occurrence}, where detection capabilities are taken into consideration. Occurrence rates are commonly measured in exoplanet science \citep[e.g.][]{2011ApJ...738..151C,2015ApJ...809....8B,2020A&A...641A.170S,2021A&A...653A.114S,2023MNRAS.521.3663B}. Here, we estimate pulsator occurrence from the measured pulsator fraction by also considering pulsation amplitude and apparent magnitude/color.

To investigate the extent to which pulsator fraction is affected by incompleteness, we plot the maximum amplitude vs. Gaia $G_{\rm BP}-G_{\rm RP}$ in Figure \ref{fig:complete}, similar to \cite{2025arXiv250818589M}. The non-detections have maximum amplitudes between $\sim10$ and $100$ ppm, increasing for redder, dimmer stars. This ``non-detection ridge'' reflects the white noise levels in our light curves. In contrast, most $\delta$ Scuti stars in NGC 3532 show maximum amplitudes about an order of magnitude higher than the non-detections, with an average maximum amplitude of $\sim 860$ ppm across all $\delta$ Scuti stars. These amplitudes are consistent with those of $\delta$ Scuti stars in the Pleiades and the Cep-Her Complex. Furthermore, we do not see many $\delta$ Scuti stars in the Pleiades and Cep-Her with amplitudes similar to those of the non-detections in NGC 3532. This suggests that our $\delta$ Scuti sample is largely complete in the central part of the instability strip.

\begin{figure*}
    \centering
    \includegraphics[width=\linewidth]{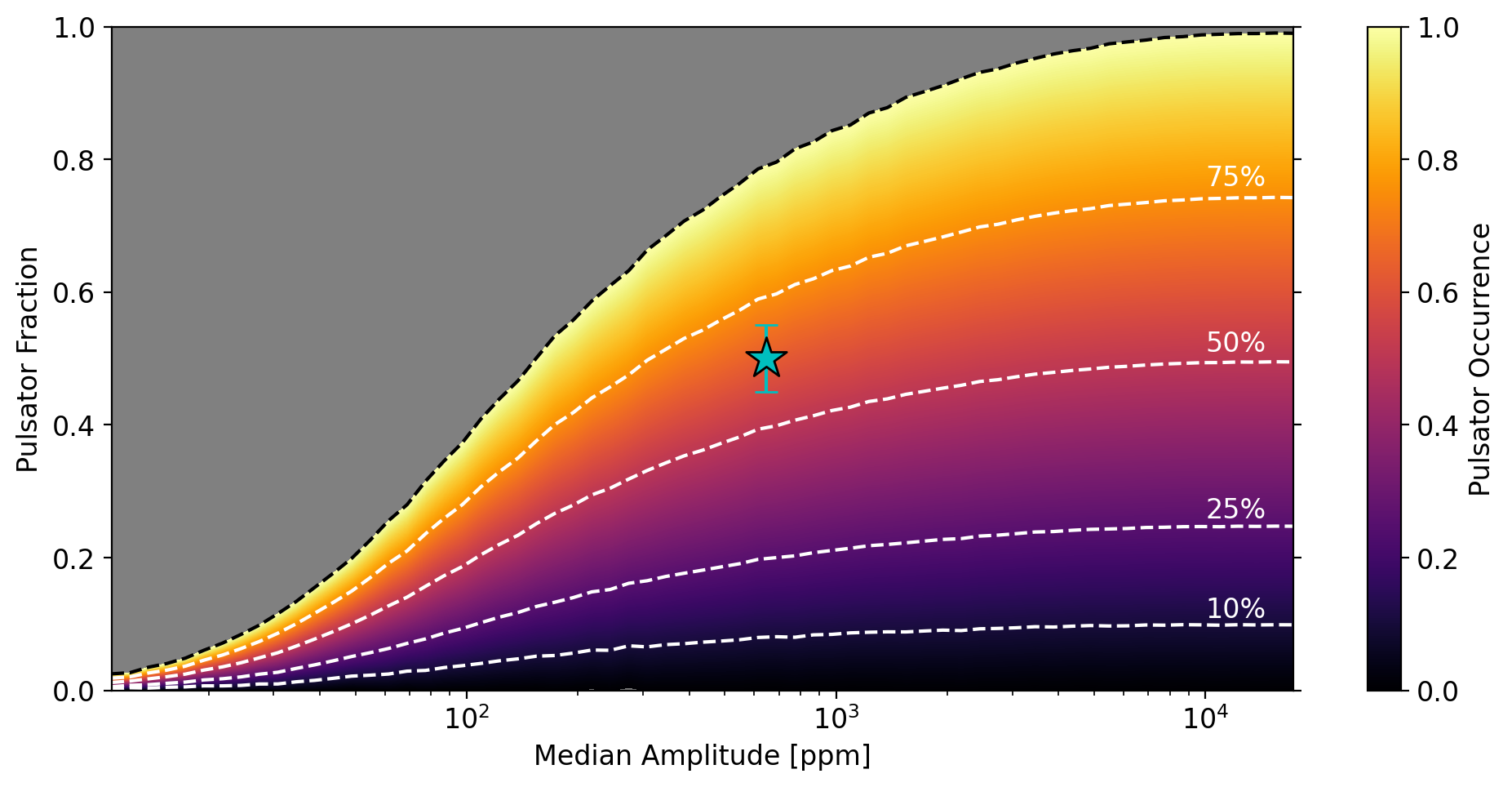}
    \caption{Pulsator occurrence (contours) as a function of pulsator fraction and median amplitude for the sample of stars in NGC 3532. Contours of constant pulsator occurrence are shown as the white dashed curves (black dashed for $100\%$). The cyan star shows the maximum pulsator fraction and median amplitude for NGC 3532.}
    \label{fig:frac_v_amp_v_occur}
\end{figure*}

As a first step to estimate pulsator occurrence, we conducted injection and recovery tests using the light curves of our non-detections. We injected a single sinusoid per test with an amplitude ranging from 1 to $10^5$ ppm \citep[based on existing empirical amplitude distributions, e.g.][]{2011MNRAS.417..591B,2019MNRAS.485.2380M,2024ApJ...972..137G} with a random frequency and phase offset. This allows us to find recovery rates as a function of maximum amplitude and apparent magnitude. A recovery occurs when the log skewness of the peak amplitudes in the resulting amplitude spectrum exceeds 0.75. The results of these injection and recovery tests are shown in Figure \ref{fig:recovery}.

Next, we conducted a Monte Carlo simulation, where each star was randomly assigned as a pulsator or non-pulsator according to its recovery fraction, determined by its actual maximum amplitude and apparent magnitude. Pulsator fraction was then measured after each star is assigned. This process was repeated 10,000 times, and we take the pulsator occurrence as the mean fraction across all trials. 

From Figure \ref{fig:pulse_frac} we see that the pulsator occurrence closely follows the pulsator fraction, but at slightly elevated levels. Specifically, we see a maximum pulsator occurrence of 67$\pm5\%$, a 7$\%$ increase from the maximum pulsator fraction. The pulsator occurrence across the center of the instability strip is 59$\pm5\%$, a 9$\%$ increase from the pulsator fraction. These results show that even if we did miss some $\delta$ Scuti stars due to detection limits, the resulting difference in the measured pulsator fraction would only be about $10\%$. Furthermore, Figure \ref{fig:recovery} shows that we can find all $\delta$ Scuti stars in this cluster with amplitudes down to $\sim500$ ppm. Two-thirds of the $\delta$ Scuti stars pulsate with amplitudes above 500 ppm 

As an alternative method to measure pulsator occurrence, we conducted a Monte Carlo simulation in which NCG 3532 stars were randomly assigned a pulsation amplitude from the \citet{2019MNRAS.485.2380M}
\textit{Kepler} distribution in Figure \ref{fig:amp_dists}. For a given pulsator occurrence, the number of recovered pulsators  was found as the sum of recovery rates based on each star's apparent magnitude and assigned amplitude. The pulsator fraction is then the number of recoveries divided by the total sample size. This process was repeated for pulsator occurrences between zero and one, and for varying median amplitudes by shifting the \textit{Kepler} distribution in log-space.

The results in Figure \ref{fig:frac_v_amp_v_occur} for this second method show that pulsator occurrence can be inferred for any population of $\delta$ Scuti stars by measuring pulsator fraction and median pulsation amplitude, assuming an underlying amplitude distribution. Across the center of the instability strip, the observed pulsator fraction and median amplitude imply a pulsator occurrence of 63$\pm6\%$, in excellent agreement with our previous estimate. These results suggest that about $37\%$ of stars in the center of the instability strip do not pulsate, and the pulsator fraction underestimates the true occurrence by $\sim13\%$. We adopt $63\%$ as the pulsator occurrence over the earlier $67\%$ estimate, as it incorporates both recovery rates and an underlying amplitude distribution.

If the pulsator occurrence were truly $100\%$ given the $50\%$ pulsator fraction, then the median amplitude would have to be $\sim 150$ ppm, roughly a factor of 4 smaller. If we treated all stars in the center of the instability strip as pulsators, then the median amplitude is $\sim 200$ ppm, which corresponds to an 88$\%$ pulsator occurrence.

\begin{figure}
    \centering
    \includegraphics[width=\columnwidth]{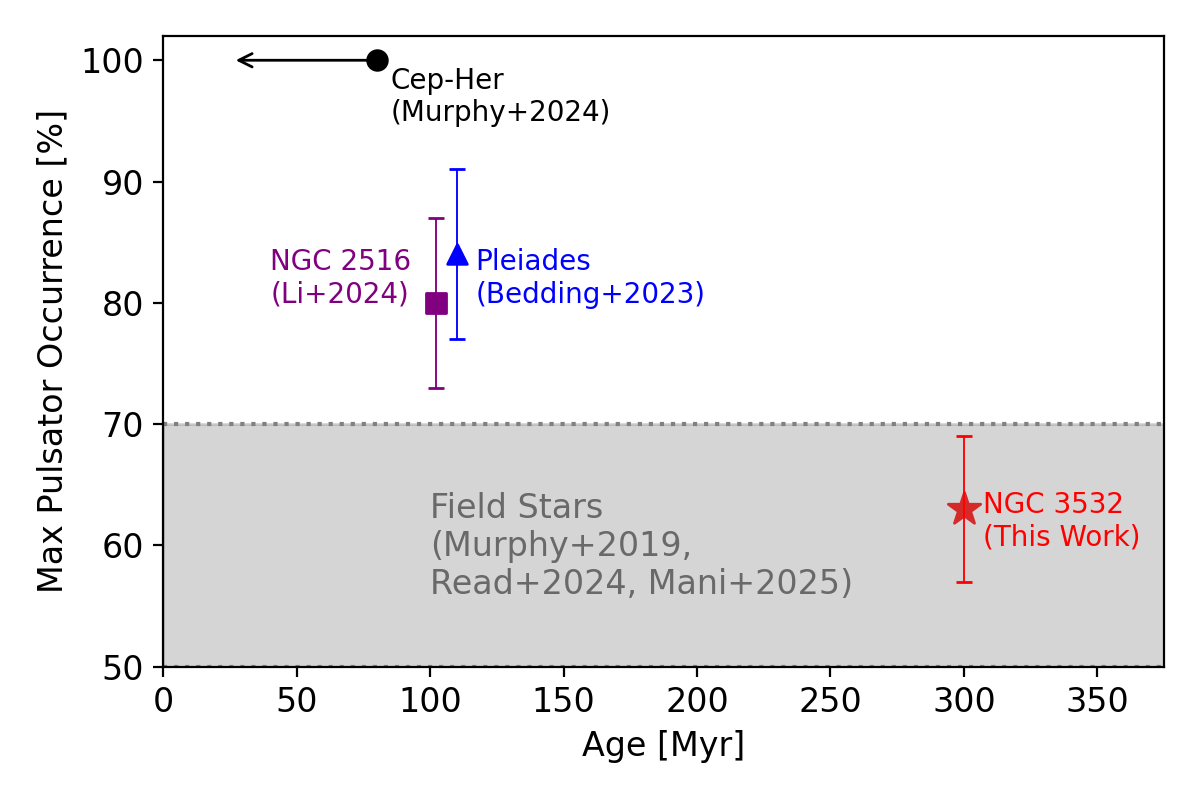}
    \caption{Maximum pulsator occurrence vs. Age for the Cep-Her Complex \citep{2024MNRAS.534.3022M}, NGC 2516 \citep{2024A&A...686A.142L}, the Pleiades \citep{2023ApJ...946L..10B}, NGC 3532 (This work), and field stars \citep{2019MNRAS.485.2380M,2024MNRAS.528.2464R, 2025arXiv250818589M}. Cep-Her, NGC 2516, and the Pleiades show pulsator fraction measurements, but the occurrence corrections are likely small.}
    \label{fig:frac_pop}
\end{figure}

Figure \ref{fig:frac_pop} shows the maximum pulsator occurrence as a function of age, indicating that younger populations show higher pulsator fractions than older populations. The likely physical explanation involves helium diffusion, which has been shown to operate on timescales similar to the cluster ages \citep[80\% depletion by 500 Myr in a 1.6${\rm M_\odot}$ star]{2005A&A...443..627T,2016A&A...589A.140D}. In young stellar populations, not enough time has passed for enough helium to sink out of the ionization zone, meaning that even slow rotators should still pulsate, and the pulsator fraction will be high. As time passes, slower rotators may stop pulsating, while rapid rotators continue to pulsate, as rotational mixing keeps helium within the ionization zone \citep{2004A&A...425..591H}. This is consistent with the trends seen in Figure \ref{fig:frac_pop}.

\section{The Period-Luminosity Relation} \label{sec:P_L}

\begin{figure*}
    \centering
    \includegraphics[width=\linewidth]{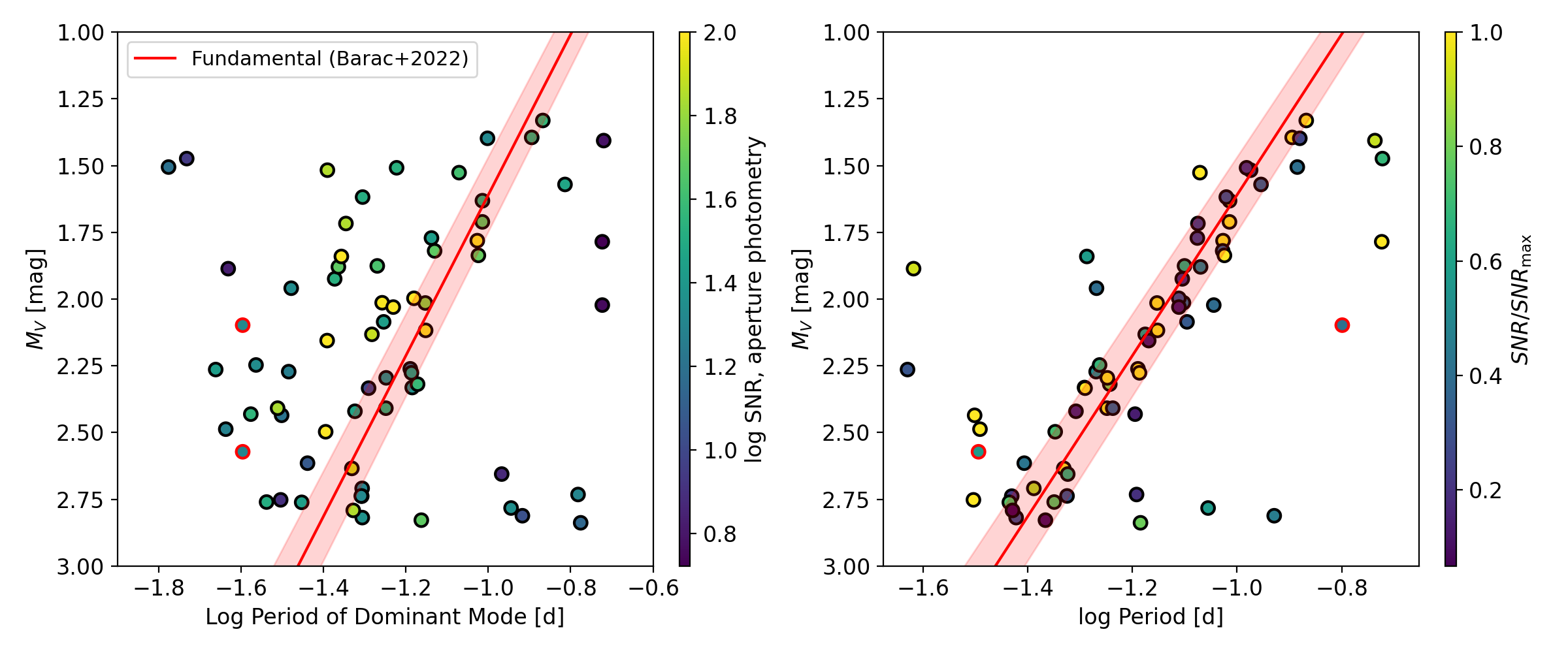}
    \caption{\textit{Left}: Period-Luminosity (P-L) relation for $\delta$ Scuti stars in NGC 3532. Colors show the SNR of the dominant mode. The fundamental mode P-L relation and uncertainty from \cite{2022MNRAS.516.2080B} is shown in red. Points with red edges are blended $\delta$ Scuti stars (see \S\ref{subsec:blend}). \textit{Right}: Same as the left panel but using the \cite{2022MNRAS.516.2080B} P-L relation to search for modes that could be the fundamental mode. Colors here denote the ratio SNR of the chosen mode to the SNR of the dominant mode.}
    \label{fig:P_L}
\end{figure*}

The left panel of Figure \ref{fig:P_L} shows the period--luminosity (P--L) relation of $\delta$ Scuti stars in NGC 3532.  We observe a large amount of scatter around the empirical P-L relation from \cite{2022MNRAS.516.2080B} (henceforth B22), which is widely assumed to follow the frequency of the fundamental radial mode \citep[$n=1, \ell=0$;][] {2000PASP..112.1096M}.

B22 found that pulsators in which the dominant mode is the fundamental show higher amplitudes than those whose dominant mode is an overtone (i.e. a pulsation mode with $n>1$). We do not observe that trend, as stars with high amplitudes (SNR$\approx$100) appear both on and off the B22 relation. 

We note that the B22 sample consists of $\delta$ Scuti stars that were discovered from ground-based observations, and which therefore have higher amplitudes than most of the stars in our sample. It seems likely that the fundamental mode is more likely to be the dominant mode in higher-amplitude $\delta$ Scuti stars.
Indeed, some of the stars in the B22 are also known high-amplitude $\delta$ Scuti stars (HADS), which pulsate in the fundamental and first overtone radial modes, with amplitudes $>0.3$ mag \citep{1999A&A...352..547P,2000PASP..112.1096M,2004CoAst.145...42R}. None of the $\delta$ Scuti stars we find in NGC 3532 are HADS.

Some stars in NGC 3532 fall close to the B22 relation, suggesting that the dominant pulsation mode is the fundamental radial mode. However, most stars fall to the left of the B22 line (at higher frequencies), most likely because their dominant mode is an overtone. This is not unexpected, however, as many $\delta$ Scuti stars in the B22 sample also fall above and to the left of the fundamental mode ridge (see their Figure 2), which they attribute to the third or fourth overtones \citep[see also][]{2019MNRAS.486.4348Z,2020MNRAS.493.4186J,2024MNRAS.528.2464R, 2025arXiv250818589M}. 

To find stars that are possibly pulsating in the fundamental radial mode, even if it is not their dominant mode, we identified peaks using iterative sine-wave fitting with the \texttt{MultiModes} Python program \citep{2022MNRAS.513..374P}, and found the highest amplitude mode within the bounds of the B22 P-L relation. If no mode was found within the B22 bounds, then we used the mode closest to the expected frequency. We only considered modes with SNR$>6$ if using aperture photometry, or SNR$>5$ if using PSF photometry. These results are shown in the right panel of Figure \ref{fig:P_L}. Of the 79 $\delta$ Scuti stars in NGC 3532, 44 show a significant pulsation mode with a frequency consistent with the expected fundamental mode frequency from B22. Only in 13 of these stars is this apparent fundamental mode also the dominant mode. We emphasize that the P-L relation is only approximate, so stars falling outside of the bounds of the B22 relation could still be fundamental mode pulsators.

Some stars show pulsation modes with frequencies to the right of the B22 relation. We set a lower frequency bound at 5 ${\rm d^{-1}}$, which is commonly used as the bound to distinguish between high-frequency $\delta$ Scuti pulsations and other sources of low frequency variability \citep[e.g. $\gamma$ Dor pulsations;][]{Kaye_1999}. The existence of hybrid $\gamma$ Dor/$\delta$ Scuti pulsators is now well established \citep{2010AN....331..989G, 2011MNRAS.417..591B, 2014MNRAS.444..102K,2015MNRAS.447.3264S,2016A&A...592A.116S,2019MNRAS.490.4040A,2022A&A...666A.142S,2024A&A...688A..25S}, so some of the stars in this regime could be hybrids with dominant g-mode pulsations, some of which cross the 5 ${\rm d^{-1}}$ boundary. A possible example in NGC 3532 is TIC 306845207, which shows apparent high amplitude $\gamma$ Dor pulsations past 10 ${\rm d^{-1}}$, but also shows possible low SNR $\delta$ Scuti pulsations up to 30 ${\rm d^{-1}}$. 

\section{The $\nu_{\rm max}$--$T_{\rm eff}$ Relation} \label{sec:vmax_Teff}

The frequency of maximum power ($\nu_{\rm max}$) is commonly used for solar-like oscillators and defined as the peak of the oscillation power excess. It has been suggested to scale as $\nu_{\rm max}\propto g/\sqrt{T_{\rm eff}}$ \citep{1991ApJ...368..599B,1995A&A...293...87K,2011A&A...530A.142B}, where $g$ is the surface gravity. 
If the $\nu_{\rm max}$ scaling relation for solar-like oscillators holds for $\delta$ Scuti pulsators, then we would expect to see slower pulsations in hotter stars, since $g$ remains relatively constant across the instability strip.

We calculate $\nu_{\rm max}$ as the amplitude-weighted mean pulsation frequency (referred to as the ``moment''), using mode frequencies identified from \texttt{MultiModes} \citep{2022MNRAS.513..374P}. Other methods, including \textit{power} weighting, or convolving the amplitude spectrum with a wide Gaussian, typically provide similar $\nu_{\rm max}$ values \citep{2024MNRAS.534.3022M}.

\begin{figure}
    \centering
    \includegraphics[width=\columnwidth]{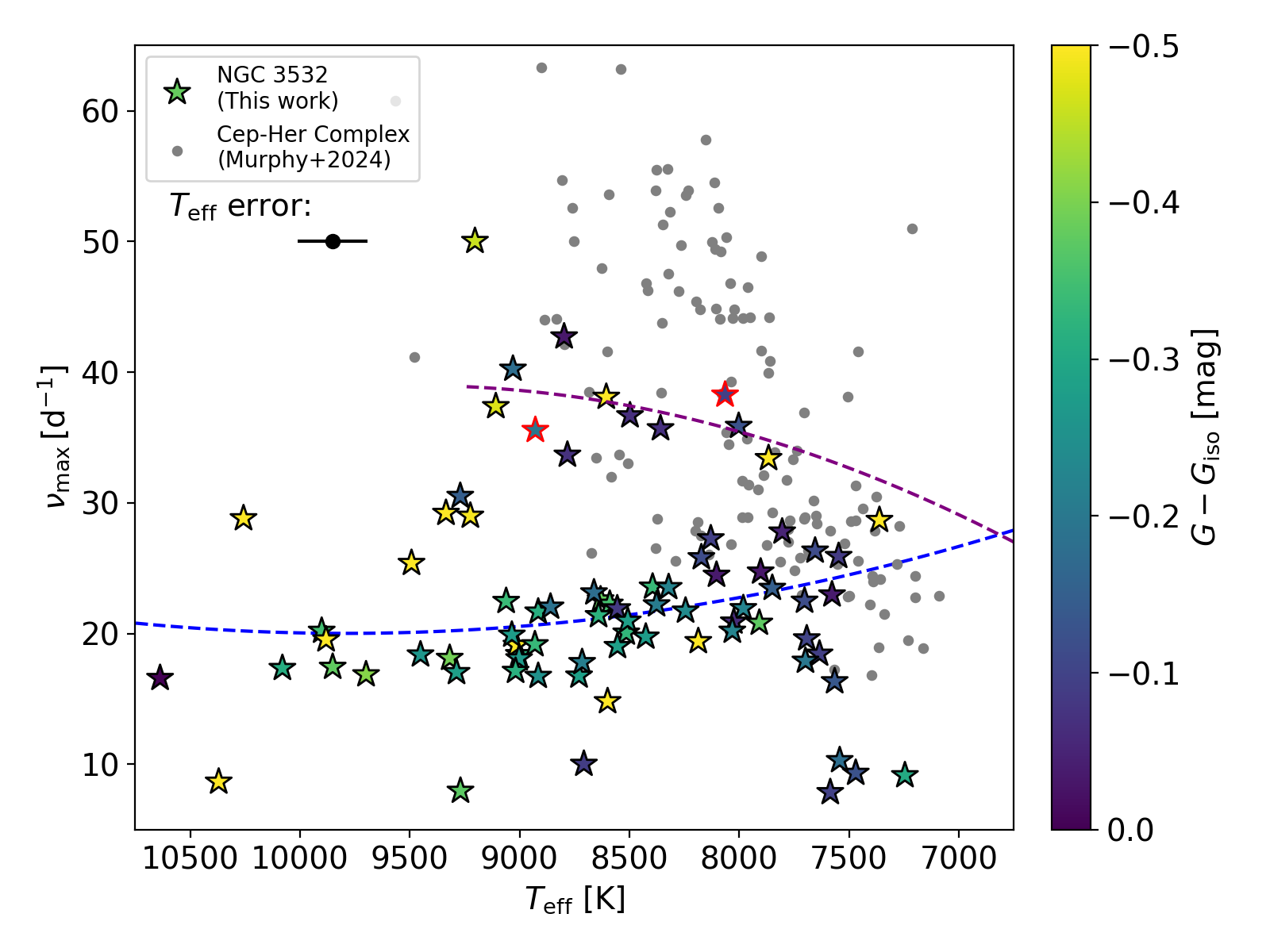}
    \caption{$\nu_{\rm max}$--$T_{\rm eff}$ relation for $\delta$ Scuti stars in NGC 3532 (color-mapped points) and Cep-Her Complex \citep[gray points;][]{2024MNRAS.534.3022M}. Colors show the vertical distance from the isochrone fit. Stars with red edges are blended $\delta$ Scuti stars (see \S\ref{subsec:blend}). Two main branches can be see in the $\nu_{\rm max}$--$T_{\rm eff}$ relation in NGC 3532, and we plot two quadratic functions for each branch to help guide the eye.}
    \label{fig:vmax_Teff}
\end{figure}

Figure \ref{fig:vmax_Teff} shows the $\nu_{\rm max}$--$T_{\rm eff}$ relation for $\delta$ Scuti stars in NGC 3532, along with results from the Cep-Her Complex \citep{2024MNRAS.534.3022M}. We see two distinct branches, characterized by higher and lower frequency pulsators, which we fit with two quadratic functions to help guide the eye. The presence of an upper branch and a lower branch is similar to the $\nu_{\rm max}$--$T_{\rm eff}$ relation in the Cep-Her Complex, and \cite{2024MNRAS.534.3022M} found that the stars on the lower branch were rotating more rapidly than stars on the upper branch. This separation is related to the discovery of two ridges in the P--L relation of $\delta$ Scuti stars that are separated by a factor of two in frequency \citep[see][]{2019MNRAS.486.4348Z,2020MNRAS.493.4186J,2024MNRAS.528.2464R, 2025arXiv250818589M}.

Generally, higher frequency pulsators are present in the Cep-Her Complex than in NGC 3532. This can be attributed to the age differences between the two populations (25-80 Myr vs. 300 Myr). On the main sequence, younger stars are expected to pulsate at higher frequencies than older stars \citep{2020Natur.581..147B,2023MNRAS.526.3779M}. 

\section{Measurement of $\Delta\nu$} \label{sec:Dnu}

The large frequency separation ($\Delta\nu$) is the mean spacing between modes of the same angular degree $\ell$ and consecutive radial order $n$.
This parameter is especially useful for mode identification, which can then be used to infer fundamental stellar properties such as age \citep[e.g.][]{2020Natur.581..147B,2021MNRAS.502.1633M,2022A&A...664A..32S,2022ApJ...941...49K,2022ApJ...941..143K,2022MNRAS.511.5718M,2023MNRAS.526.3779M,2023MNRAS.525.5235S}.
However, measuring $\Delta\nu$ is not always feasible, as it requires modes to be regularly spaced in frequency. This is most commonly seen in young ($\lesssim$ 100 Myr) $\delta$ Scuti stars that pulsate in high radial order $n\gtrsim 5$ \citep{2020Natur.581..147B}. Furthermore, \cite{2020Natur.581..147B} and \cite{2024MNRAS.534.3022M} found that many $\delta$ Scuti stars with regular pulsations have low $v\sin i \leq 50$ km/s, suggesting that rapid rotation spoils regular mode spacings.

To measure $\Delta\nu$ for $\delta$ Scuti stars in NGC 3532 we followed \cite {2022MNRAS.513..374P,2023A&A...675A.167P} (and references within), who used the Fast Fourier Transform (FFT), Autocorrelation Function (ACF), and the Histogram of Frequency Differences (HFD) on the peaks in the amplitude spectrum. They also used the \'echelle diagram \citep{2022ascl.soft07005H} to verify $\Delta\nu$.

\begin{figure}
    \centering
    \includegraphics[width=\columnwidth]{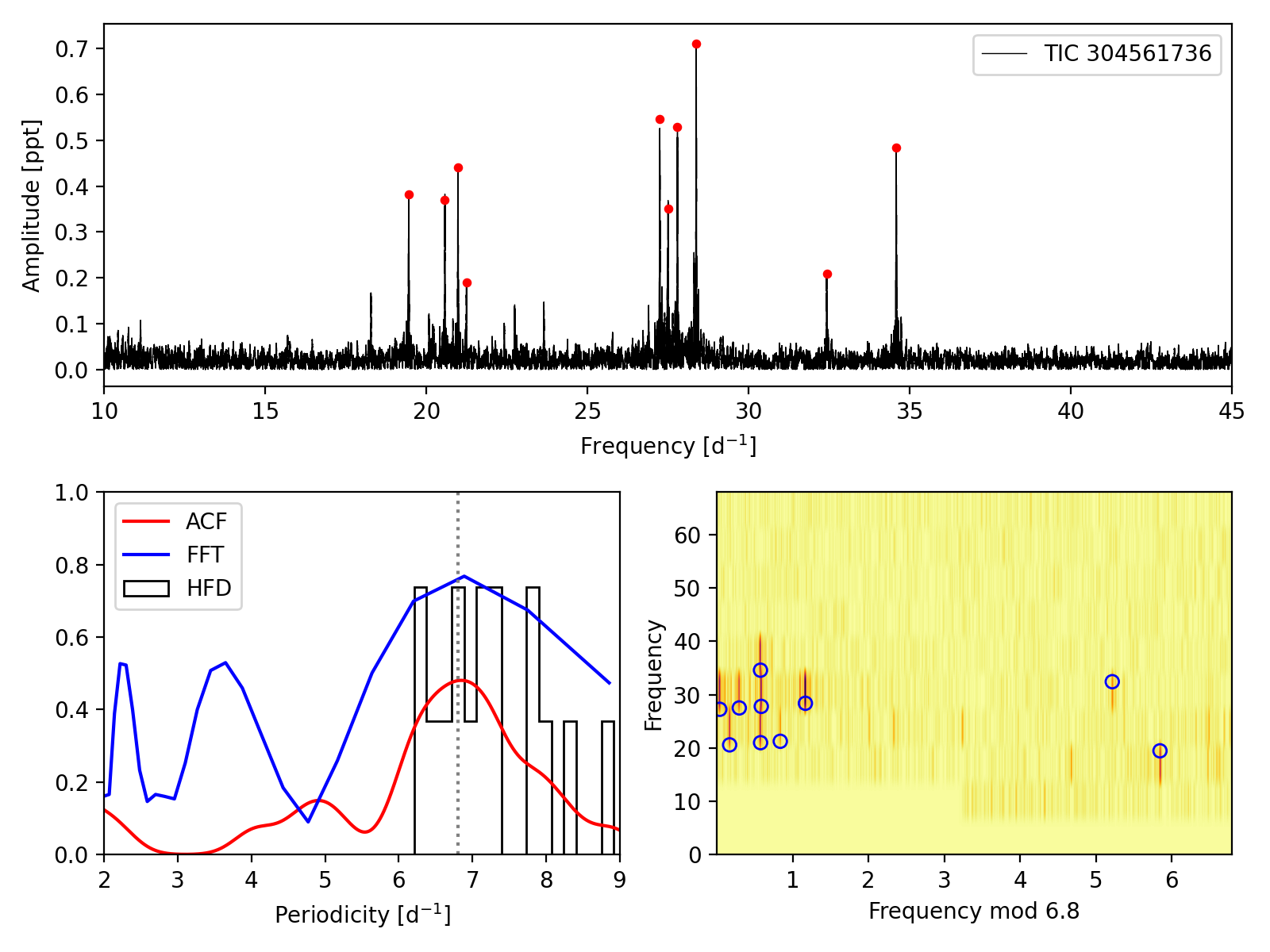}
    \caption{\textit{Top}: Amplitude spectrum for TIC 304561736 with the 10 highest pulsation modes used for $\Delta\nu$ measurement marked with red points. \textit{Bottom Left}: The Autocorrelation Function (ACF, red), Fast Fourier Transform (FFT, blue) and Histogram of Frequency Differences (HFD, black). The gray dotted line shows the measured $\Delta\nu=6.8\,{\rm d^{-1}}$. \textit{Bottom Right}: \'Echelle diagram using the found $\Delta\nu$. The blue circles show the locations of the ten marked modes in the amplitude spectrum.}
    \label{fig:dnu}
\end{figure}

We successfully measured $\Delta \nu$ for only one $\delta$ Scuti in NGC 3532, TIC 304561736. We find $\Delta\nu=6.8\,{\rm d^{-1}}$ (see Figure \ref{fig:dnu}), which is consistent for a $\delta$ Scuti star of this age \citep{2020Natur.581..147B}. The amplitude spectrum is characterized by tight groups of modes spaced by $\Delta\nu$. Based on the \'echelle diagram  the lowest frequency mode could be the fundamental mode, and the next three modes just above 20\,${\rm d^{-1}}$ could be a $\ell=1$ rotationally split triplet. The next group of modes near $\approx28\,{\rm {d^{-1}}}$ are perhaps the first overtone ($n=2$) and the corresponding $\ell=1$ triplet. The next mode at $\approx32\,{\rm {d^{-1}}}$ is likely the second radial overtone ($n=3$) and the highest frequency mode should be the corresponding $\ell=1$ mode, though it does not appear to be split into a triplet. This star has a low \texttt{vbroad}$\approx20$ km/s, which may be consistent with split modes close to the $\ell=1,m=0$ modes. The low number of stars with regular spacings is consistent with the fact that the cluster is relatively old and has many rapid rotators. 

\section{Conclusions} \label{sec:Conclusion}

In this work we have reported the discovery of 79 $\delta$ Scuti stars in the 300 Myr old open cluster NGC 3532. This is the most $\delta$ Scuti stars found in a individual open cluster to-date, over a factor of two more than the Pleiades \citep{2023ApJ...946L..10B}. We then analyzed these $\delta$ Scuti stars and obtained the following results:

\begin{enumerate}
    \item We found 79 $\delta$ Scuti pulsators in NGC 3532, which are rotating on average a factor of two faster than their non-pulsating counterparts. This shows that rotation is highly important in maintaining $\delta$ Scuti pulsations over the main sequence lifetime.
    \item We calculated the pulsator fraction and introduced the concept of pulsator occurrence, which takes into account the detection capabilities of TESS and an underlying amplitude distribution. In the 300 Myr old cluster NGC 3532, we found a  pulsator fraction of 50$\pm5\%$ and a pulsator occurrence of 63$\pm6\%$. This is lower than pulsator fractions found in the Pleiades \citep[84\%, 110 Myr;][]{2023ApJ...946L..10B}, NGC 2516 \citep[$\sim 80\%$, 100 Myr;][]{2024A&A...686A.142L}, and the Cep-Her Complex \citep[100\%, $\leq$ 80 Myr;][]{2024MNRAS.534.3022M}. This suggests that $\delta$ Scuti stars can stop pulsating over time.
    \item We investigated the P--L relation and found much scatter around the expected fundamental P--L relation from \citep{2022MNRAS.516.2080B}. This suggests the dominant mode of the $\delta$ Scuti stars in NGC 3532 could be a higher radial or non-radial overtone mode.
    \item We found the $\nu_{\rm max}$--$T_{\rm eff}$ relation in NGC 3532 and found two distinct branches, similar to that of the Cep-Her Complex. Overall we see lower $\nu_{\rm max}$ in NGC 3532 than Cep-Her, which could be attributed to the older age of NGC 3532.
    \item We measured $\Delta\nu$ for one $\delta$ Scuti in NGC 3532, 6.8 ${\rm d^{-1}}$.
\end{enumerate}

Much work remains in the study of $\delta$ Scuti pulsators, both in NGC 3532 and in other open clusters. Some $\delta$ Scuti stars in NGC 3532 may have been missed in this work, as some possible members may have been overlooked due to poor astrometric solutions from Gaia DR3 \citep[though most stars with poor astrometric solutions have Gaia \textit{G}$>$14 mag;][]{2025arXiv250414672T}. Many A and F-type stars in NGC 3532 lack \texttt{vbroad} measurements. Spectroscopy should be obtained for these and other cluster members to measure $v\sin i$ and test the results shown in Figure \ref{fig:dGiso}. In addition, rotational modulation could be searched for in these stars, which would provide a rotation period and a viewer inclination. Additional stars may also yield measurable $\Delta\nu$ with more careful analysis, enabling mode identification and a seismically inferred cluster age. Finally, similar studies can be conducted in many other open clusters, such as the 200 Myr old open cluster Messier 7.

\section*{Acknowledgements}
I.B. and D.H. acknowledge support from the National Aeronautics and Space Administration (80NSSC22K0781). T.R.B and S.J.M. were supported by the Australian Research Council through Laureate Fellowship FL220100117 and Future Fellowship FT2100100485.

This paper includes data collected with the TESS mission, obtained from the MAST data archive at the Space Telescope Science Institute (STScI). Funding for the TESS mission is provided by the NASA Explorer Program. STScI is operated by the Association of Universities for Research in Astronomy, Inc., under NASA contract NAS 5–26555.

\section*{Data Availability Statement}
The data presented in this paper are available via a machine-readable table. This work used data from the TIC catalog \citep{TIC_catalog}, and FFI photometry from TESS Sectors 37, 63, and 64 \citep{sector_37,sector_63,sector_64}. The TGLC light curves generated for this work are available at DOI: \dataset[10.5281/zenodo.17402160]{https://doi.org/10.5281/zenodo.17402160}.

\begin{deluxetable*}{cccccccccccc}
\tablecaption{First 50 entries from our NGC 3532 target catalog; the full version with additional columns is available in machine-readable format. Columns list TIC ID, apparent Gaia $G$, right ascension ($\alpha$), declination ($\delta$),  $G_{\rm BP}-G_{\rm RP}$, $G-G_{\rm iso}$, $T_{\rm eff}$, luminosity ($L$), \texttt{vbroad}, maximum amplitude ($A_{\rm max}$), frequency of maximum power ($\nu_{\rm max}$), and a $\delta$ Scuti flag (1 for detections, 0 for non-detections). \label{tab:delta_scuti}}
\tablehead{
\colhead{TIC} & \colhead{Gaia $G$} & \colhead{$\alpha$ [$^\circ$]} & \colhead{$\delta$ [$^\circ$]} & \colhead{$G_{\rm BP}-G_{\rm RP}$} & \colhead{$G-G_{\rm iso}$} &
\colhead{$T_{\rm eff}$ [K]} & \colhead{$L$ [$L_\odot$]} & 
\colhead{\texttt{vbroad} [km/s]}  & \colhead{$A_{\rm max}$ [ppm]} & \colhead{$\nu_{\rm max}$ [${\rm d^{-1}}$]} & \colhead{$\delta$ Scuti}
}
\startdata
307744790 & 10.23 & 167.25 & -59.49 & 0.19 & -0.18 & 8666 & 19 & \nodata & 58.29 & \nodata & 0 \\
308413664 & 8.37 & 167.57 & -59.51 & 0.11 & -1.65 & 9464 & 116 & 178 & 39.66 & \nodata & 0 \\
308947425 & 8.03 & 167.92 & -59.39 & 0.03 & 0.62 & 10216 & 183 & 79 & 21.78 & \nodata & 0 \\
307746783 & 10.18 & 167.37 & -59.27 & 0.13 & -0.02 & 9285 & 23 & \nodata & 54.50 & \nodata & 0 \\
306919558 & 10.47 & 167.10 & -59.30 & 0.23 & -0.12 & 8583 & 17 & \nodata & 50.51 & \nodata & 0 \\
308282115 & 10.64 & 167.50 & -59.08 & 0.30 & -0.13 & 8098 & 13 & \nodata & 65.86 & \nodata & 0 \\
307748562 & 10.13 & 167.37 & -59.07 & 0.21 & -0.38 & 8631 & 22 & \nodata & 443.42 & 22.60 & 1 \\
306917505 & 11.13 & 167.14 & -59.05 & 0.47 & -0.30 & 7245 & 6 & 19 & 461.97 & 9.15 & 1 \\
923933908 & 11.22 & 167.14 & -58.95 & 0.48 & -0.24 & 7182 & 6 & \nodata & 5171.23 & \nodata & 0 \\
305447938 & 11.38 & 165.88 & -59.54 & 0.50 & -0.12 & 7120 & 5 & 59 & 113.77 & \nodata & 0 \\
305447284 & 10.72 & 165.88 & -59.46 & 0.32 & -0.13 & 8002 & 11 & 203 & 919.51 & 35.88 & 1 \\
304992327 & 11.14 & 165.59 & -59.35 & 0.43 & -0.09 & 7634 & 8 & 129 & 600.66 & 18.45 & 1 \\
304467732 & 9.85 & 165.21 & -59.18 & 0.17 & -0.47 & 9108 & 33 & \nodata & 207.09 & 37.39 & 1 \\
304468490 & 11.31 & 165.30 & -59.09 & 0.45 & -0.01 & 7487 & 6 & \nodata & 96.90 & \nodata & 0 \\
304365599 & 7.95 & 165.10 & -58.98 & 0.08 & -1.82 & 10370 & 219 & 175 & 1036.83 & 8.66 & 1 \\
306496114 & 10.73 & 166.76 & -59.38 & 0.35 & -0.22 & 7834 & 11 & 32 & 73.05 & \nodata & 0 \\
306390348 & 9.99 & 166.73 & -59.41 & 0.16 & -0.28 & 8952 & 23 & \nodata & 3700.13 & \nodata & 0 \\
306389586 & 10.29 & 166.63 & -59.31 & 0.22 & -0.30 & 8585 & 17 & \nodata & 67.13 & \nodata & 0 \\
306388620 & 10.65 & 166.68 & -59.19 & 0.29 & -0.04 & 8200 & 11 & 74 & 99.85 & \nodata & 0 \\
306043574 & 10.33 & 166.51 & -59.21 & 0.21 & -0.10 & 8756 & 19 & \nodata & 198.21 & \nodata & 0 \\
306044327 & 11.18 & 166.47 & -59.12 & 0.46 & -0.13 & 7469 & 7 & 203 & 408.47 & 9.34 & 1 \\
305912454 & 9.10 & 166.31 & -59.08 & 0.04 & 1.39 & 11068 & 96 & \nodata & 33.38 & \nodata & 0 \\
305912066 & 10.22 & 166.34 & -59.04 & 0.20 & -0.18 & 9030 & 21 & 295 & 161.50 & 40.25 & 1 \\
305458263 & 10.68 & 166.13 & -59.15 & 0.50 & -0.74 & 7356 & 11 & \nodata & 130.84 & \nodata & 0 \\
305911904 & 9.01 & 166.28 & -59.02 & 0.04 & -0.53 & 10974 & 103 & \nodata & 136.45 & \nodata & 0 \\
305911920 & 7.73 & 166.24 & -59.02 & 0.04 & -1.82 & 10853 & 290 & 135 & 24.72 & \nodata & 0 \\
305911543 & 10.88 & 166.19 & -58.97 & 0.38 & -0.09 & 7987 & 10 & \nodata & 183.10 & \nodata & 0 \\
305911591 & 10.39 & 166.21 & -58.98 & 0.25 & -0.28 & 8729 & 19 & 176 & 1654.58 & 16.77 & 1 \\
306498425 & 9.61 & 166.85 & -59.10 & 0.13 & -0.50 & 9269 & 37 & 315 & 39.80 & \nodata & 0 \\
306387638 & 9.38 & 166.67 & -59.08 & 0.09 & -0.52 & 9670 & 35 & \nodata & 38.76 & \nodata & 0 \\
306917605 & 10.25 & 166.98 & -59.06 & 0.36 & -0.82 & 7738 & 18 & 50 & 49.11 & \nodata & 0 \\
306917111 & 9.58 & 167.02 & -59.00 & 0.06 & -0.14 & 9756 & 42 & \nodata & 30.23 & \nodata & 0 \\
306499684 & 9.88 & 166.96 & -58.95 & 0.15 & -0.32 & 9003 & 28 & 90 & 806.87 & 18.42 & 1 \\
306500448 & 10.66 & 166.95 & -58.86 & 0.27 & 0.00 & 8158 & 12 & \nodata & 58.63 & \nodata & 0 \\
306499552 & 10.29 & 166.83 & -58.97 & 0.23 & -0.27 & 8425 & 17 & 94 & 460.34 & 19.75 & 1 \\
306500067 & 8.29 & 166.80 & -58.91 & 0.01 & 0.14 & 10293 & 142 & 23 & 19.78 & \nodata & 0 \\
306385801 & 9.52 & 166.66 & -58.86 & 0.04 & 1.76 & 10120 & 44 & 9 & 154.51 & \nodata & 0 \\
306385684 & 11.05 & 166.64 & -58.84 & 0.40 & -0.09 & 7547 & 7 & 103 & 510.54 & 25.92 & 1 \\
306500803 & 9.61 & 166.79 & -58.82 & 0.13 & -0.59 & 9039 & 37 & 278 & 44.58 & \nodata & 0 \\
306500891 & 10.79 & 166.84 & -58.80 & 0.36 & -0.21 & 7666 & 10 & \nodata & 97.30 & \nodata & 0 \\
306385513 & 10.09 & 166.74 & -58.82 & 0.19 & -0.35 & 8589 & 18 & 307 & 858.25 & 22.27 & 1 \\
306385358 & 10.16 & 166.68 & -58.80 & 0.21 & -0.33 & 8514 & 18 & \nodata & 2571.83 & 20.06 & 1 \\
306385286 & 7.37 & 166.73 & -58.79 & 0.07 & -2.35 & 9645 & \nodata & 179 & 19.76 & \nodata & 0 \\
306385106 & 9.72 & 166.73 & -58.77 & 0.12 & -0.37 & 9269 & 34 & \nodata & 698.91 & 7.96 & 1 \\
306384924 & 8.69 & 166.71 & -58.74 & 0.04 & 1.06 & 10108 & 94 & \nodata & 31.88 & \nodata & 0 \\
306045026 & 10.34 & 166.54 & -59.04 & 0.18 & -0.01 & 8989 & 17 & \nodata & 55.91 & \nodata & 0 \\
306045270 & 11.31 & 166.45 & -59.01 & 0.45 & -0.00 & 7470 & 6 & \nodata & 105.39 & \nodata & 0 \\
305911595 & 10.93 & 166.34 & -58.98 & 0.36 & -0.05 & 7947 & 9 & 78 & 67.87 & \nodata & 0 \\
306045628 & 10.10 & 166.43 & -58.96 & 0.13 & -0.06 & 9477 & 25 & \nodata & 51.90 & \nodata & 0 \\
306045824 & 11.10 & 166.54 & -58.94 & 0.40 & -0.04 & 7577 & 6 & 101 & 856.64 & 22.98 & 1 \\
\bottomrule
\enddata
\end{deluxetable*}

\clearpage

\bibliographystyle{aasjournal}
\bibliography{export-bibtex.bib}
\clearpage

\end{document}